\documentclass{aa}

\usepackage{graphicx}
\usepackage{txfonts}

\usepackage[]{amsmath}     
\usepackage{natbib}        
\bibpunct{(}{)}{;}{a}{}{,} 
\usepackage{color}

\begin{document}

\title{Forming the cores of giant planets \\ 
from the radial pebble flux in protoplanetary discs }
\titlerunning{Forming the cores of giant planets}
\author{
M. Lambrechts \and A. Johansen
} 

\institute{Lund Observatory, Department of Astronomy and Theoretical
Physics, Lund University, Box 43, 22100 Lund, Sweden\\
\email{michiel@astro.lu.se}
}
   
\date{Received --- ; accepted ---}

\abstract{
The formation of planetary cores must proceed rapidly in order for the giant
planets to accrete their gaseous envelopes before the dissipation of the
protoplanetary gas disc ($\lesssim 3$\,Myr).  
In orbits beyond $10$\,AU, direct accumulation of planetesimals by the cores is
too slow. 
Fragments of planetesimals could be accreted faster, but planetesimals are
likely too large for fragmentation to be efficient, and resonant trapping poses
an additional hurdle.  
Here we instead investigate the accretion of small pebbles (mm-cm sizes) that
are the natural outcome of an equilibrium between the growth and radial drift
of particles. 
We construct a simplified analytical model of dust coagulation and pebble drift
in the outer disc, between $5$\,AU and $100$\,AU, which gives the
temporal evolution of the solid surface density and the dominant particle size. 
These two key quantities determine how core growth proceeds at various orbital
distances.
We find that pebble surface densities are sufficiently high to achieve the
inside-out formation of planetary cores within the disc lifetime. The overall
efficiency by which dust gets converted to planets can be high, close to 50\%
for planetary architectures similar to the Solar System.
Growth by pebble accretion in the outer disc is sufficiently fast to
overcome catastrophic Type I migration of the cores. 
These results require protoplanetary discs with large radial extent
($\gtrsim$$100$\,AU) and assume a low number of initial seed embryos.
Our findings imply that protoplanetary discs with low disc masses, as expected
around low-mass stars ($<1$\,M$_\odot$), or with sub-solar dust-to-gas ratios,
do not easily form gas-giant planets ($M\gtrsim 100$\,M$_{\rm E}$), but
preferentially form Neptune-mass planets or smaller ($M\lesssim 10$\,M$_{\rm
E}$). 
This is consistent with exoplanet surveys which show that gas giants are
relatively uncommon around stars of low mass or low metallicity.
}
   

\maketitle

%
%

\section{Introduction}

The giant planets in our Solar System have a large fraction of heavy elements in
their interiors \citep{Guillot_2005}. 
Models of giant exoplanets in short orbits show similar core masses, on the
order of $10$ Earth masses (M$_{\rm E}$) or larger \citep{Guillot_2006,
Miller_2011, Moutou_2013}.
In the core accretion scenario, the formation of these cores is the critical
first step in a process of attracting a gaseous envelope \citep[][]{Pollack_1996}.

However, core formation faces two hurdles. 
Firstly, growth by pairwise collisions comes to a halt when particles reach
radii between mm and cm. 
Gas drag on these pebbles leads to a rapid radial migration on time scales of
$100$-$1000$\,yr \citep{Weidenschilling_1977}. 
This prevents continued particle growth by coagulation \citep{Brauer_2008},
except possibly in the inner disc with large particle seeding
\citep{Windmark_2012} or porous dust aggregates \citep{Kataoka_2013}. 
Therefore the migration of solids leads to a decrease in the solid surface in
the outer disc and stops particle growth beyond pebble-size.

Secondly, even if solids would overcome the radial drift barrier and reach
sizes larger than kilometer (`planetesimals'), collisional growth towards
core-sizes remains too slow. 
Observationally, gas discs are estimated to survive only for a few Myr
\citep{Haisch_2001,Kraus_2012}. 
Models which assume that core growth occurs by the accretion of planetesimals
larger than km in size \citep{Pollack_1996, Kobayashi_2011} cannot form
the cores of the giant planets at distances of $\sim$$5$\,AU or larger within
this time scale, without evoking significantly enhanced solid surface densities
compared to the Minimum Mass Solar Nebula \citep[MMSN,][]{Hayashi_1981}.
Destruction of large planetesimals into smaller fragments can speed up
core growth \citep{Rafikov_2004, Ormel_2010}. 
However, in order to collide sufficiently frequently planetesimals have to be
small \citep[$\sim$1\,km, ][]{Chambers_2014}, below current estimates
\citep[$\sim$$100$\,km, ][]{Morbidelli_2009}, and global disc simulations show
that growth is hampered by trapping large fragments in resonances
\citep{Levison_2010}.

These two issues -- the loss of solids by pebble drift and slow planetesimal
accretion -- can be overcome by considering the accretion of pebbles onto larger
planetesimals and cores.
Firstly, numerical simulations find very high accretion rates, which allows core
formation well within disc lifetimes \citep{Lambrechts_2012}.
This is due to gas drag operating on pebbles that enter the
gravitational reach of a core, resulting in kinetic energy dissipation and
accretion \citep{Johansen_2010,Ormel_2010}. 
Larger planetesimals would need much closer encounters to be gravitationally focussed
onto the core.
Secondly, a large fraction of drifting pebbles can be accreted by the growing
cores \citep{Morbidelli_2012}, therefore avoiding the loss of pebbles to
sublimation at evaporation fronts interior to the giant planet formation zone.

The evolution of the solid surface density and the dominant pebble size on the
scale of the protoplanetary disc are critical parameters in determining core
growth rates and accretion efficiencies. The aim of this paper is to explore
pebble accretion in global models of particle growth and pebble drift.
We introduce a simplified model of dust growth that relies on separating
the population of stationary dust grains from larger drifting pebbles
(Section\,2). 
We verify that this model is in agreement with more detailed numerical
coagulation codes and observational constraints.
The growth of planetary cores is subsequently explored at various orbital
distances, based on parametrized pebble accretion rates from hydrodynamical
simulations (Section\,3).
In this regime of fast pebble accretion, we find that Type I migration does not
lead cores into the star, a result which validates our approximation of in situ
growth (Section\,4).
Planetary architectures like our Solar System can form efficiently in an
inside-out fashion, under standardly assumed gas column densities and
dust-to-gas ratios. However, variations in these parameters can radically
change the mass of the planets that form in these systems (Section\,5).
Finally, we conclude the paper with highlighting some areas for future
exploration (Section\,6) and a brief summary of our results (Section\,7).

\section{The evolution of the solid surface density}
\subsection{The protoplanetary disc}

We start by setting up a model for the gas component of the protoplanetary
disc. 
Although complex prescriptions can in principle be included in our model, we
choose a simple prescription of the gas surface density as
\begin{align}
\Sigma_{\rm g} = \beta \left( \frac{r}{\rm AU} \right)^{-1},
\end{align}
where we can either choose $\beta = \beta_0$ a fixed normalisation constant or
$\beta = \beta_0 \exp(-t/\tau_{\rm dis})$ in order to mimic gas dissipation, as
done for example by \citet{McNeil_2005} and \citet{Walsh_2011}.
This radial power law profile is expected from the viscous evolution of an
accretion disc \citep{Lynden_1974} and supported by the observed disc profile
of the nearby protoplanetary disc around the star TW Hya \citep{Andrews_2012}. 
For simplicity we will use here $\tau_{\rm dis} = 3$\,Myr \citep{Haisch_2001},
but disc lifetimes could be even shorter for stars above a solar mass
\citep{Hernandez_2005}.

The thermal profile of the disc ($T,c_{\rm s}, H/r$) follows the standard MMSN
values, which should be fairly accurate in the outer parts of the disc where
viscous heating is negligible \citep[][]{Bitsch_2013, Bitsch_2014}, and we will find that our
results only depend weakly on these choices.

\subsection{Dust growth}

We employ a simple, but robust, model for particle growth which takes into
account the vertical settling of particles in the disc, 
inspired by the approach by \citet{Garaud_2007} and \citet{Birnstiel_2012}.
The growth rate of a particle can be expressed as 
\begin{align}
  \dot R &= \frac{1}{4} \frac{\rho_{0,d}}{\rho_\bullet} \Delta v_{\rm t}\,,
  \label{eq:turb_sweep}
\end{align}
when a particle of size $R$ and material density $\rho_\bullet$ sweeps up
particles of sizes smaller than $R$ in a midplane layer of particle density $\rho_{0,d}$. 
The relative velocity, $\Delta v_{\rm t}$, between particles is driven by the
turbulent gas motions. In the so-called intermediate regime
\citep{Ormel_2007}, the collision velocities are parametrized by
\begin{align}
  \Delta v_{\rm t} = \sqrt{3\alpha_{\rm t} \tau_{\rm f}}c_{\rm s}, 
  \label{eq:turb_v}
\end{align}
with $c_{\rm s}$ the sound speed, $\alpha_{\rm t}$ the turbulent viscosity
parameter and $\tau_{\rm f}$ the Stokes number (the product of the friction time
of the particle and the Keplerian frequency $\Omega_{\rm K}$).
Since the vertical scale height of particles depends on the strength of the
turbulence in the disc through 
\begin{align}
  H_{\rm d} \approx H \sqrt{\alpha_{\rm
  t}/\tau_{\rm f}}
  \label{eq:par_scale}
\end{align}
\citep{Youdin_2007}, the growth rate becomes independent of
$\alpha_{\rm t}$, 
\begin{align}
  \dot R = \frac{\sqrt{3}}{4} \frac{\rho_0}{\rho_\bullet} Z c_{\rm s}
  \tau_{\rm f}
  \label{eq:dotR}
\end{align}
\citep{Brauer_2008}.
Here $Z$ is the \emph{local} ratio of the dust and gas column densities
and $\rho_0$ is the midplane gas density. 
The Stokes number $\tau_{\rm f}$ is connected to the particle size via
\begin{align}
  \tau_{\rm f} = \frac{\rho_\bullet R}{\rho_0 c_{\rm s}} \Omega_{\rm K}.
\end{align}
Here we have assumed the Epstein drag regime, which is generally valid in the
giant planet formation zone where the gas mean free path is larger than
pebble sizes.
Therefore the growth time scale does not depend on the internal density
(porosity) of the particles,
\begin{align}
  t_{\rm g} =\frac{R}{\dot R} 
  = \frac{4}{\sqrt{3} \epsilon_{\rm g} Z \Omega_{\rm K}} \,.
  \label{eq:grow_timescale_turb}
\end{align}
Here we have introduced $\epsilon_{\rm g}$, a parameter that can be
used to modify the growth efficiency between particles. 

In our calculation we have ignored growth driven by Brownian motion and radial
and azimuthal differential drift, which are small contributions in turbulent
discs \citep{Brauer_2008}. Our analysis also assumes that the sticking
coefficient $\epsilon_{\rm g}$ is independent of particle size.
This is certainly not the case, especially for larger particles, for
which relative velocities become large. However for small dust grains, perfect
sticking is a relatively good approximation. For larger mm-sized particles,
collision outcomes predominantly lead to fragmentation \citep{Blum_2008},
but collision outcomes between icy particles are not well known. With the help of the free parameter $\epsilon_{\rm g}$ we can explore the
coagulation efficiency.

\subsection{Dust mass flux}

The outer parts of the protoplanetary discs act as reservoir of solid
particles \citep{Youdin_2002,Garaud_2007}.
At wide orbits small stationary dust slowly grows into inwards-drifting
pebbles, which in turn drive pebble accretion in the zone where the giant
planets form.
At an orbital radius $r$, the characteristic time scale $t_{\rm g,d}$ for dust
to grow is given by Eq.\,(\ref{eq:grow_timescale_turb}), with $Z$ taken to be
the initial dust-to-gas ratio $Z_0$ and $\epsilon_{\rm g}$
taken to be the dust sticking efficiency $\epsilon_{\rm g,d}$. 
We choose a standard dust-to-gas ratio of $Z_0=0.01$ \citep{Draine_2007},
unless mentioned otherwise.
For the sticking efficiency we take $\epsilon_{\rm g,d}=0.5$, which produces
results consistent with more advanced coagulation codes \citep{Birnstiel_2012}.
The particle-growth time scale is independent of particle size
(Eq.\,\ref{eq:grow_timescale_turb}).
However, the actual time used to grow from ISM-like dust with radius $R_0$ to
the size at which pebbles start drifting $R_{\rm drift}$ is given by
\begin{align}
  \Delta t = \ln(R_{\rm drift}/R_0) t_{\rm grow,d} 
  \approx  \xi  t_{\rm grow,d},
  \label{eq:rat}
\end{align}
with a weak logarithmic dependence on the size ratio $R_{\rm drift}/R_0$.
The size of $R_{\rm drift}$ is not known a priori, but we find that $R_{\rm
drift} \approx 1-10$\,mm is a characteristic size and consistent with
observations. 
ISM grains are predominantly about $0.1-1 \mu$m in size
\citep{Weingartner_2001}, which would result in $\xi \approx 10$.
For simplicity, we now write 
\begin{align}
  \Delta t  &= \frac{4}{\sqrt{3} \epsilon_{\rm d} Z_0 \Omega_{\rm K}},
  \label{eq:simplified}
\end{align}
with $\epsilon_{\rm d} = \epsilon_{\rm g,d} \xi^{-1}$ encapsulating the two particle
growth parameters.

After a time $\Delta t$ of exponential growth, a particle will reach a size
where the growth rate and drift rate become comparable, and the pebble starts
drifting towards the star.
We can now find, for a certain time $t$, where in the disc particles
have just grown to pebble sizes (by setting $t = \Delta t$ in
Eq.\,\ref{eq:simplified}), 
\begin{align}
  r_{\rm g}(t) = 
  \left( \frac{3}{16} \right)^{1/3}
  (GM_*)^{1/3} 
  \left( \epsilon_{\rm d} Z_0\right)^{2/3} 
  t^{2/3}.
\end{align}
Here $G$ is the gravitational constant and $M_*$ the stellar mass.
This expression has previously been found to be in good agreement with more
detailed numerical coagulation models \citep{Garaud_2007}.
The pebble production line, $r_{\rm g}(t)$, moves outwards and sweeps up a pebble mass flux
\begin{align}
  \dot M_{\mathcal F} = 
  2\pi r_{\rm g} \frac{dr_{\rm g}}{dt} \Sigma_{d,0}(r_{\rm g}).
  \label{eq:mdot}
\end{align}
Here, the derivative $dr_{\rm g}/dt$ denotes the speed at which the pebble
front expands radially,
\begin{align}
  \frac{dr_{\rm g}}{dt} = 
  \frac{2}{3} \left( \frac{3}{16} \right)^{1/3}
  (GM_*)^{1/3} 
  \left(\epsilon_{\rm d} Z_0 \right)^{2/3} 
  t^{-1/3}.
\end{align}
When we assume that the initial dust surface density is a constant fraction of
the gas surface density,
\begin{align}
  \Sigma_{d,0} 
  = Z_0\, \beta \left( \frac{r}{\rm AU} \right)^{-1},
  \label{eq:Zassum}
\end{align}
the radial pebble flux through the protoplanetary disc can be expressed as
\begin{align}
  \dot M_{\mathcal F} =& (2/3)^{2/3} \pi 
  (GM_*)^{1/3} 
  (\beta\,{\rm AU}) \epsilon_{\rm d}^{2/3} Z_0^{5/3}
  t^{-1/3} \nonumber \\
  \approx& 9.5 \times 10^{-5} 
  \left( \frac{\beta}{500\,{\rm g\,cm}^{-2}} \right)  \
  \left( \frac{M_*}{\rm M_\odot} \right)^{1/3} \nonumber\\
  &\times \left( \frac{Z_0}{0.01} \right)^{5/3}
  \left( \frac{t}{10^6\,{\rm yr}} \right)^{-1/3} 
  {\rm M}_{\rm E}\,{\rm yr}^{-1}\,,
  \label{eq:theMdot}
\end{align}
showing only a weak time dependence. Here $M_\odot$ is the solar mass.
Although, as expected, the radial pebble flux depends on the initial dust-to-gas
ratio and the coagulation efficiency, this expression only depends on the disc
structure through the gas surface density, but not on the temperature.

\subsection{Evolution of the solid surface density}

\begin{figure}
  \includegraphics{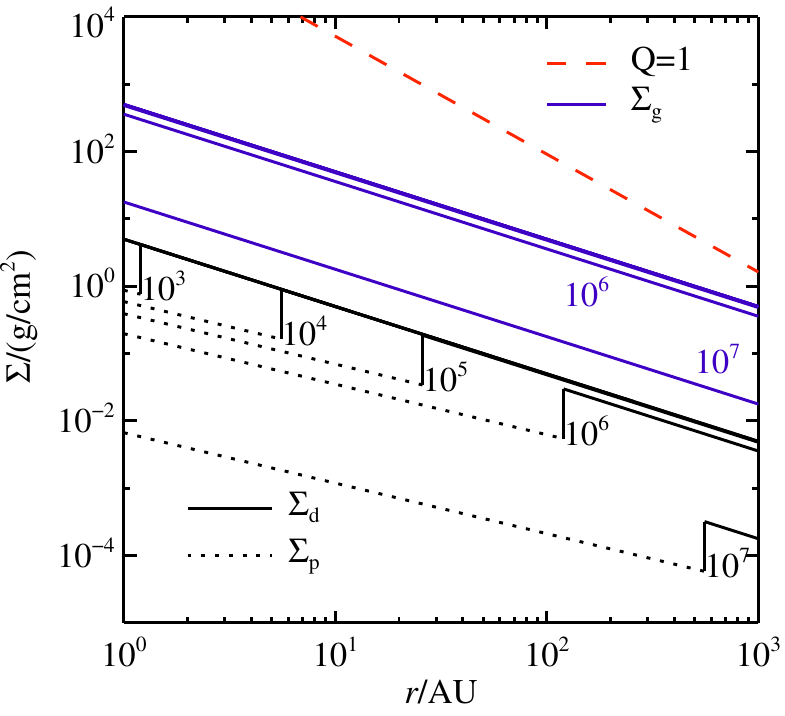}
  \caption{
  Solid surface density as function of orbital distance (black), at different
  times (indicated by the labels representing the time in years). The
  solid black lines represent the dust surface density and the dotted lines the
  pebble surface density. 
  The exponentially decaying gas surface density is represented in blue, for
  the same times as the solid surface density. 
  We have here assumed a gas disc dissipation time scale of $\tau_{\rm dis} =
  3$\,Myr.
  Gas surface densities larger than the red dashed line (which represent Toomre
  $Q=1$), would be gravitationally unstable (see also Sec.\,\ref{sec:Mdisc}).
  }
  \label{fig:sig}
\end{figure}

Once the pebble mass flux is set by the production of pebbles in the outer
disc, we can calculate the surface density interior to $r_{\rm g}$. 
From the continuity requirement, we find 
\begin{align}
  \dot M_{\mathcal F} = 2\pi r v_{ r} \Sigma_{\rm p},
  \label{eq:simple}
\end{align}
where 
$\Sigma_{\rm p}(r)$ is the pebble surface density interior to this outer 
reservoir.
In this expression, the particle velocity $v_{r}$ depends on the pebble
size, which is also a function of $\Sigma_{\rm p}$. 
The particle size can be estimated from balancing the growth time scale with the
drift timescale. 
Following Eq.\,(\ref{eq:grow_timescale_turb}), we find the growth time scale
for drifting pebbles to be
\begin{align}
  t_{\rm g, p} = \frac{4}{\sqrt{3} \epsilon_{\rm p} (\Sigma_{\rm
  p}/\Sigma_{\rm g}) \Omega_{\rm K}}\,,
  \label{eq:pebble_growth}
\end{align}
with $\epsilon_{\rm p}$ the coagulation efficiency between pebbles (we will
assume $\epsilon_{\rm p} =0.5$, similar to the dust coagulation efficiency
$\epsilon_{\rm g,d}$).

Particles drift radially inwards with a velocity
\begin{align}
   v_r = - 2 \frac{\tau_{\rm f}}{\tau_{\rm f}^2+1} \eta v_{\rm K}
  \label{eq:v_r}
\end{align}
\citep{Weidenschilling_1977, Nakagawa_1986}, 
where $v_{\rm K}$ is the Keplerian velocity at orbital radius $r$ and $\eta$ is
a dimensionless measure of radial gas pressure support,
\begin{align}
  \eta = -\frac{1}{2} \left( \frac{H}{r} \right)^2
           \frac{\partial \ln P}{\partial \ln r} 
       = 0.0015 \left( \frac{r}{\rm AU} \right)^{1/2}.
  \label{eq:eta_param}
\end{align}
The turbulent motion of the pebbles in the radial direction can be safely
ignored for the purpose of determining the bulk radial drift\footnote{
The average radial particle velocity is given to good approximation by
Eq.\,(\ref{eq:v_r}), even when considering turbulence and collective particle
effects \citep{Weidenschilling_2006}. 
Because the diffusive timescale over a length $l$ is given by $t_{\rm diff} =
l^2/(\alpha_{\rm t} H^2 \Omega_{\rm K})$ and the radial drift time scale by
$t_{\rm drift} =
l/(2\tau_{\rm f} \eta v_{\rm K})$, diffusion only dominates on scales smaller than 
$(l/H) \approx \alpha (H/r) / (2\tau_{\rm f} \eta)
\approx 0.062 (\alpha_{\rm t}/10^{-3})(\tau_{\rm f}/0.1)^{-1} (r/{\rm
10\,AU})^{-1/4}$. This is much smaller than
the global scale of the disc, $r$, over which pebble transport occurs.
}.
In the limit of a particle with $\tau_{\rm f} \lesssim 1$, we then find a radial drift time scale 
\begin{align}
  t_r = \frac{r}{v_r} 
      \approx 550 \left( \frac{\tau_{\rm f}}{0.1} \right)^{-1}
      \left( \frac{r}{\rm AU} \right)\,{\rm yr.} 
  \label{eq:drift_ts}
\end{align}
We now obtain, by setting $t_{\rm g,p} = t_r$, the dominant particle size of
\begin{align}
  \tau_{\rm f} &\approx \frac{\sqrt{3}}{8}  
  \frac{\epsilon_{\rm p}}{\eta}
  \frac{\Sigma_{\rm p}}{\Sigma_{\rm g}}
  \label{eq:max_st}
\end{align}
\citep{Brauer_2008, Birnstiel_2012}.
Equivalently, the above expression for the size of pebbles can also be found by
combining Eq.\,(\ref{eq:dotR}) and Eq.\,(\ref{eq:v_r}), resulting in 
\begin{align}
  \frac{d\tau_{\rm f}}{dr} = \frac{\dot \tau_{\rm f}}{v_r} 
  = - \frac{\sqrt{3}}{8}  
  \frac{ \epsilon_{\rm p}}{\eta} 
  \frac{\Sigma_{\rm p}}{\Sigma_{\rm g}} \frac{1}{r}\,.
  \label{eq:full_deriv}
\end{align}
After integration we find 
\begin{align}
  \tau_{\rm f} =- \frac{\sqrt{3} \epsilon_{\rm p}}{8} \frac{\xi_0}{\beta_0 \eta_0}
  \frac{1}{\psi+1/2}
  \left[ \left( \frac{r}{\rm AU} \right)^{\psi+1/2}
  \right]^{r/{\rm AU}}_{r_0/{\rm AU}} +\tau_{\rm f,0} \,,
  \label{eq:full_deriv2}
\end{align}
where we have assumed the pebble surface density can be described as a power
law function of the form $\Sigma_{\rm p} = \xi_0 (r/{\rm AU})^{\psi}$, with
$\psi <-1/2$ (which we verify later to be valid). Similarly, we used $\eta =
\eta_0 (r/{\rm AU})^{1/2}$.
The particles are of size $\tau_{\rm f,0}$ at initial location $r_0$, but at $r
\ll r_0$ their size is approximately given by 
\begin{align}
  \tau_{\rm f} \approx - \frac{\sqrt{3}}{8} \frac{1}{\psi+1/2} \frac{\epsilon_{\rm
  p}}{\eta} \frac{\Sigma_{\rm p}}{\Sigma_{\rm g}}\,,
  \label{eq:full_deriv3}
\end{align}
comparable to Eq.\,\ref{eq:max_st}.

Therefore, knowing the pebble size, we can rewrite Eq.\,(\ref{eq:simple}) as
\begin{align}
  \Sigma_{\rm p} = \sqrt{\frac{2 \dot M_{\mathcal F} \Sigma_g}{\sqrt{3} \pi
  \epsilon_{\rm p} r v_{\rm K}}}\,.
  \label{eq:pebblesurfacedens}
\end{align}
Notice that the pebble surface density no longer depends on the pressure profile
and gas disc scale height through $\eta$.
The same expression is found in \citet{Birnstiel_2012}, although we can here   
combine it with our analytical expression for the pebble flux                   
$\dot{M}_{\mathcal F}$ from Eq.\,(\ref{eq:theMdot}).
In this way we obtain the temporal and radial dependency of the pebble column
densities,
\begin{align}
  \Sigma_{\rm p}(r,t) =& 
  2^{5/6} 3^{-7/12} 
  \frac{\epsilon_{\rm d}^{1/3}}{\epsilon_{\rm p}^{1/2}}
  Z_0^{5/6} \Sigma_{\rm g} \Omega_{\rm K}^{-1/6} t^{-1/6} \nonumber\\
  \approx& 0.069
  \left( \frac{\beta}{500 {\rm g/cm}^2} \right)
  \left( \frac{Z_0}{0.01} \right)^{5/6}
  \left( \frac{M_*}{ {\rm M}_\odot} \right)^{-1/12}
  \left( \frac{t}{10^6\,{\rm yr}} \right)^{-1/6}   \nonumber\\
  &\times \left( \frac{r}{10\,{\rm AU}} \right)^{-3/4}
  \,{\rm g\,cm}^{-2}\,,
  \label{eq:full}
\end{align}
which is valid in the region $r < r_{\rm g}$.
The value of the pebble surface density should be quite robust, as it does not
depend on the temperature structure of the disc through $H/r$. Also, there
is only a weak dependency on the coagulation efficiency through  the ratio
$\epsilon_{\rm d}^{1/3} / \epsilon_{\rm p}^{1/2}$.

Figure\,\ref{fig:sig} illustrates the gas and solid surface densities as
function of orbital radius and time.
While the dust surface density inherits the radial profile of the gas surface
density ($\Sigma_{\rm d} \propto r^{-1}$), the slope of the pebble surface
density falls off as $r^{-3/4}$. 
In principle, one could interpolate between the two regimes, but an
instantaneous transition is sufficient for our purposes.
Similarly, we have not included an exponential edge to the disc here, but this
mainly affects the disc beyond $\approx$$100$\,AU \citep{Birnstiel_2012}.

The power law profile for the pebble column density is based here on a
simplified model for pebble formation and  is very different compared to often
used MMSN estimates, which yield $\Sigma_{\rm p}=0.5(r/10\,{\rm
AU})^{-1.5}$\,g/cm$^2$.
However, advanced coagulation codes modelling compact particle growth yield
very similar results as the analytic calculation \citep{Brauer_2008,Birnstiel_2012, Okuzumi_2012}. 
In particular, Fig.\,3 of \citet{Okuzumi_2012} is quantitatively similar to our
Fig.\,\ref{fig:sig} (but note that the authors used a gas profile $\Sigma_{\rm
g} \propto r^{-3/2}$ as in the MMSN, which results in $\Sigma_{\rm p} \propto
r^{-1}$).
Intriguingly, a radial slope of $-3/4$ in the pebble surface density has been
inferred for the outer regions of the protoplanetary disc around TW Hydra
\citep{Andrews_2012, Birnstiel_2012, Menu_2014}.

\subsection{Dominant pebble size}

\begin{figure}
  \includegraphics{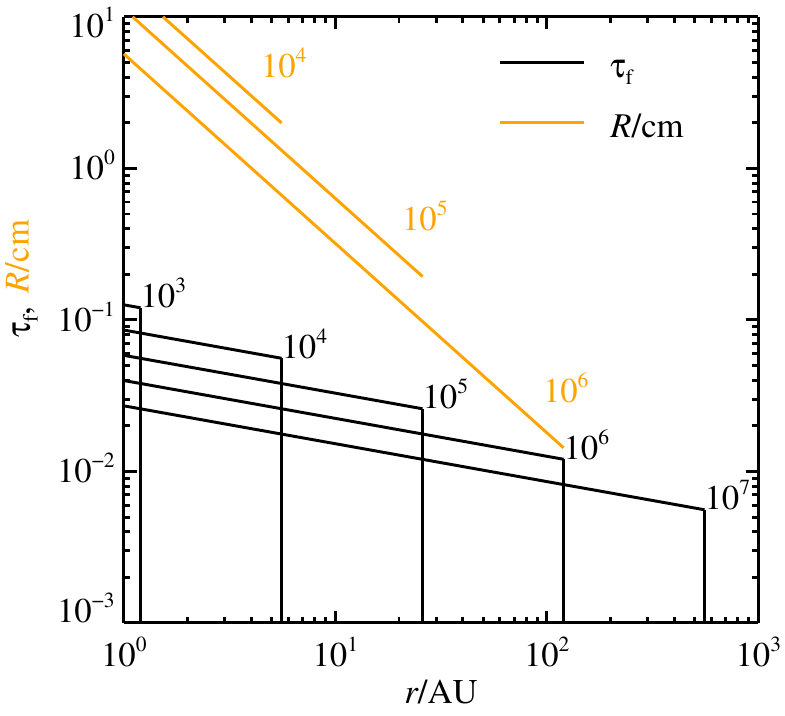}
  \caption{
  The dominant particle size, as function of orbital distance and time. 
  Black lines represent the particle size expressed by the Stokes number, the
  dimensionless friction time of the particle (labels correspond to the
  disc age in yr). 
  Yellow lines give the particle size in cm assuming the Epstein regime
  ($10^3$ and $10^7$\,yr omitted for clarity). 
  Particle sizes decrease at wider orbits and at later times, but typically
  remain in the mm-cm regime in the giant planet formation zone (5-50\,AU). 
  Pebbles are spread to the orbital location where dust is still growing to
  pebble sizes $r_{\rm g}$. Beyond this orbit most of the solid mass will be in
  smaller dust particles. The disc dissipation time scale was set to
  $\tau_{\rm dis}=3$\,Myr.
  }
  \label{fig:Rpeb}
\end{figure}

Figure\,\ref{fig:Rpeb} shows the dominant pebble size as function of semi-major
axis and time, based on Eq.\,(\ref{eq:max_st}) and Eq.\,(\ref{eq:full}).
The pebble size is about $\tau_{\rm f} \approx 0.02$-$0.04$
around $t=1$\,Myr, in good agreement with results from advanced coagulation
codes \citep{Brauer_2008,Birnstiel_2012}.
Such particle sizes, between $1$\,cm at $5$\,AU and $1$\,mm at $30$\,AU are in
agreement with observations of the spectral slope of the dust emission from
protoplanetary discs at mm to cm wavelengths.
A large fraction of young discs contain a significant dust mass fraction in
solids of mm sizes which remain present over the lifetime of
discs \citep{Draine_2006, Natta_2007,Ricci_2010,Ubach_2012}.  
In particular, observations in cm wavelengths of the nearby young star TW
Hya\footnote{The age of TW Hya is estimated to be 5-10\,Myr, but the star is
possibly as young as $\sim$$3$\,Myr based on
near-infrared spectroscopy by \citet{Vacca_2011}.} 
have revealed that as much as $\sim$$10^{-3}$\,M$_{\odot}$ of the solid mass is
present in cm-sized pebbles \citep{Wilner_2005}, while simultaneously containing $\gtrsim 0.05$\,M$_\odot$
in gas \citep{Bergin_2013}. 
Other examples are WW Cha and GG Tau A (ages $\sim$$0.5$ and $\sim$$1.5$\,Myr),
which have very similar pebble discs, with dust mass $M_d \sim
10^{-3}$\,M$_\odot$ and outer radii $r_{\rm 0} \approx 250$\,AU
\citep{Lommen_2009,Scaife_2013}. 

As illustrated in Fig.\,\ref{fig:Rpeb}, the sizes of the pebbles are smaller
in wider orbits.
Hints of a decrease in pebble size from cm in the inner disc to sub-mm in the
outer disc ($\gtrsim 75$\,AU) have also been seen in the protoplanetary disc of
AS 209 \citep{Perez_2012}, CQ Tauri \citep{Trotta_2013} and TW Hydra
\citep{Menu_2014}.
The sharp cut-off in the particle size at the pebble production line in
Fig.\,\ref{fig:Rpeb} is an artefact of our model. In reality a smooth
connection exists between the pebbles and growing dust in wider orbits.

\subsection{Radial extent of the dust disc}
\label{sec:Mdisc}
It is well established that discs need a large radial extent ($\gtrsim 100$\,AU
) to explain mm-observations \citep{Brauer_2008}, which do not indicate any
depletion of the dust mass before gas dissipation \citep{Ricci_2010}. 
The need for a large disc to act as a mass reservoir can also be seen in
Fig.\,\ref{fig:sig}, where the pebble formation edge $r_{\rm g}$ moves out in
time.
Such large radii are not inconsistent with those inferred from
protoplanetary discs \citep{Isella_2009,Williams_2011} and debris discs \citep{Wyatt_2008}.
CO observations tracing the gas component of the disc suggest a mean outer disc
radii larger than $r_{\rm o}$$\approx$$210$\,AU, for stars with ages $\lesssim
7$\,Myr, nearly independent of the stellar spectral type \citep{Dent_2005}. 
Similarly, sub-mm/mm surveys find outer radii of $\sim$$100$\,AU,
with some protoplanetary discs much larger \citep{Kitamura_2002,Mohanty_2013}.
Protoplanetary discs are expected to have a large radial extent at late times,
because of the viscous expansion of the disc \citep{Lynden_1974}. 
The outer edge is typically characterized by a sharp exponential cut-off.
However, this solution is unreliable as it in fact violates the assumption of a 
disc where the radial pressure gradient can be ignored compared to the
central gravity of the star,
which is required for these self-similar solutions \citep{Ono_2014}.

We have verified that our large disc models do not become gravitationally
unstable anywhere in the disc \citep{Toomre_1964}, see for example
Fig.\,\ref{fig:sig} where we plot the line of $Q$$=$$1$, above which a disc
would become gravitationally unstable. 
The parameter $Q$ is defined as $Q\approx c_{\rm s}\Omega_{\rm K}/(\pi G
\Sigma)$.

\subsection{Model assumptions}

The calculation of this model for the equilibrium pebble surface density and
dominant particle size relies on two important assumptions.
Firstly, particles remain small with $\tau \lesssim 1$ and in the Epstein
regime, in both the dust growth and pebble drift regimes.
Secondly, the pebble disc is in equilibrium between growth and drift (as in Eq.\,\ref{eq:simple}).
Therefore the time scale on which the radial pebble flux changes,
$\tau_{\mathcal{F}} =
r_{\rm g}/\dot r_{\rm g} =(3/2) t$,
must be larger than the pebble-drift time scale ($t_{\rm r}$,
Eq.\,\ref{eq:drift_ts}) at $r \leq r_{\rm g}$.
This criterion is satisfied for the parameters we have chosen. If, on the
other hand, the pebble coagulation efficiency would be very small, $\epsilon_{\rm p}
\rightarrow 0$, these small and slow pebbles would lead to unphysical results
such as the pebble surface density obtaining a higher value then the original
dust density.

Additionally, we have made the approximation that the growth of pebbles occurs dominantly by collisions with other pebbles, as
opposed to dust, in the region with $r<r_{\rm g}$ (Eq.\,\ref{eq:pebble_growth}).
The ratio of pebble growth by collisions with pebbles relative to collisions
with unsedimented dust is equal to the ratio of pebble and dust midplane
densities,
\begin{align}
  \frac{\dot M_{\rm p}}{\dot M_{\rm d}} \approx \frac{\rho_{\rm 0,p}}{\rho_{\rm 0,d}}
  = \frac{\Sigma_{\rm p}}{\Sigma_{\rm d}} \frac{H}{H_{\rm p}},
  \label{eq:dustvspebble}
\end{align}
because the collision speed between two
pebbles is comparable to the collision speed between a pebble and a small dust
grain \citep{Weidenschilling_1984}.
For particles of size $\tau_{\rm f} =0.1$ and turbulence strength $\alpha_{\rm
t} =10^{-3}$, we find $H/H_{\rm p} \sim 10$ (Eq.\,\ref{eq:par_scale}). 
Therefore the contribution of dust can be safely ignored, especially since we
assume that at orbital radii where $r<r_{\rm g}$ dust is efficiently turned into
pebbles ($\Sigma_{\rm p} \gg \Sigma_{\rm d}$).
The assumption that sweep-up of dust is negligible might break down in a
strongly turbulent outer disc where small pebbles are prevented from settling
to the midplane and large amounts of dust are produced in catastrophic
collisions.

Our calculation is valid for the smooth power law disc models that are typically used.
It does however not include the possible presence of pressure bumps,
regions where locally $\eta \lesssim 0$. In such areas, the particle surface
density cannot be described by a globally smooth profile and the
particle size estimate of Eq.\,(\ref{eq:full_deriv3}) no longer holds.

Finally, one could think that our results are sensitive to the porosity of the
icy particles. However, as demonstrated by \citet{Okuzumi_2012}, in the Epstein
regime (valid in the giant planet formation zone) the growth time scale is
independent of the solid density of the particles
(Eq.\,\ref{eq:grow_timescale_turb}).

\section{Core growth}
\subsection{Accretion rate}

The presence of pebbles in the outer disc will drive the rapid growth of the
cores of the giant planets.
In the previous section we outlined how particles grow by coagulation to pebble
sizes and start drifting inwards. 
Subsequently, pebbles are rapidly accreted by large planetesimals, provided
that the planetesimal mass is above the transition mass,
\begin{align}
  M_{\rm t} 
  &\approx \sqrt{\frac{1}{3}} \frac{(\eta v_{\rm
  K})^3}{G\Omega_{\rm K}} 
  \approx 0.0069 \left( \frac{r}{\rm 5\,AU} \right)^{3/2} {\rm M}_{\rm E}\,,
  \label{eq:seed}
\end{align}
which depends cubicly on the uncertain value of $\eta(r)$
\citep{Lambrechts_2012}. Here, we adopted the value given in
Eq.\,(\ref{eq:eta_param}).

In this paper, we assume that these core seed masses have already formed, but
leave their formation and expected number for future work. 
These initial cores could be the result of streaming instabilities in the
coupled motion between pebbles and gas, leading to the formation of
planetesimals of about $\sim$ $0.1\,M_{\rm t}$ \citep{Johansen_2012}, aided by
continued growth by pebble accretion in the so-called drift branch
\citep{Lambrechts_2012}.
Interestingly, there is a lack of bodies larger than $M_{\rm t}$ in the Solar
System, which is encouraging because in our model such planetesimals
would have grown to planetary sizes.

Given the surface densities of pebbles and their sizes, we use the numerical
results from \citet{Lambrechts_2012} to find the core accretion rates. 
A core accretes pebbles with Stokes numbers $\tau_{\rm f} <0.1$ at a rate of
\begin{align}
  \dot M_{\rm c}
  &= 2 
  \left( \frac{\tau_{\rm f}}{0.1} \right)^{2/3} 
  r_{\rm H}v_{\rm H} \Sigma_{\rm p}\, .
  \label{eq:Mcdot} 
\end{align}
This corresponds to the cores accreting all pebbles with $\tau_{\rm f} \approx 0.1$ that enter
the Hill sphere, with radius
\begin{align}
  r_{\rm H} = r \left( \frac{M_{\rm c}}{3M_*} \right)^{1/3}\, ,
  \label{eq:rH}
\end{align}
at a velocity $v_{\rm H} = \Omega_{\rm K} r_{\rm H}$. 
This is the result of gas drag on the pebble operating on a time scale similar
to the crossing time of the particle past the core ($\approx$$\Omega_{\rm
K}^{-1}$). 
Therefore, gas drag on a particle entering the Hill sphere slows it down,
leading to accretion by the core.
Smaller particles are more tightly coupled to the gas, so the accretion
radius diminishes, leading to the size-dependent factor in
the accretion rate in Eq.\,(\ref{eq:Mcdot}).
Finally, we have here implicitly made the assumption that the particle scale
height is comparable to or smaller than the Hill radius, which is satisfied for
standard turbulent strengths. 
For example, the particle scale height is about $H_{\rm p}/H\approx 0.1$, for
$\alpha_{\rm t}=10^{-3}$ and $\tau_{\rm f}=0.1$ (Eq.\ref{eq:par_scale}).
This is comparable to the Hill radius for a core of $0.1$\,M$_{\rm E}$,
\begin{align}
  \frac{r_{\rm H}}{H} = 0.08 
  \left( \frac{M_{\rm c}}{0.1\,{\rm M}_{\rm E}} \right)^{1/3} 
  \left( \frac{r}{10\,{\rm AU}} \right)^{-1/4}\,,
  \label{eq:r_HH}
\end{align}
assuming $H/r = 0.06\,(r/{\rm 10 AU})^{1/4}$. Therefore, only in the short time
between the masses $M_{\rm t}$ and $0.1$\,M$_{\rm E}$ will the accretion rates be
somewhat reduced by not accreting from the full particle layer. 
Furthermore, low viscosity regions, as found in dead zone models and disc wind
models \citep{Turner_2014}, could severely reduce the particle scale height.

We can make use of our disc model to express the pebble accretion rate as
\begin{align}
  \dot M_{\rm c}  
  \approx& \, 4.8 \times 10^{-6} 
  \left( \frac{M_c}{ {\rm M}_{\rm E}} \right)^{2/3}  
  \left( \frac{Z_0}{0.01} \right)^{25/18} 
  \left( \frac{M_*}{ {\rm M}_\odot} \right)^{-11/36}
  \left( \frac{\beta}{500 {\rm g\,cm}^{-2}} \right)  \nonumber\\
  &\times \left( \frac{r}{\rm 10\,AU} \right)^{-5/12}
  \left( \frac{t}{10^6\,{\rm yr}} \right)^{-5/18}
  \, {\rm M}_{\rm E}\,{\rm yr}^{-1}\, ,
  \label{eq:Mcdot_const}
\end{align}
by inserting Eq.\,(\ref{eq:max_st}) and Eq.\,(\ref{eq:full}) into
Eq.\,(\ref{eq:Mcdot}).
The growth rates of the cores are thus only weakly dependent on time and orbital distance. 
This is encouraging because Uranus and Neptune have very similar core masses,
respectively $M_{\rm c,U}\approx 13$\,M$_{\rm E}$ and $M_{\rm c,N}\approx
15$\,M$_{\rm E}$ \citep[][]{Helled_2011}.

Only a fraction $f$ of the pebbles that drift past the core are
accreted \citep{Morbidelli_2012,Ormel_2012, Guillot_2014}. We can
express this fraction as the ratio of the accreted pebbles over the rate at
which pebbles drift radially through the disc
\begin{align}
  f = \frac{\dot M_{\rm c}}{\dot M_{\mathcal F}}.
  \label{eq:frac}
\end{align}
By combining Eq.\,(\ref{eq:simple}) and Eq.\,(\ref{eq:Mcdot}), we find 
\begin{align}
  f &= \frac{5}{\pi} \left( \frac{\tau_{f}}{0.1} \right)^{-1/3} 
  \eta^{-1} \left( \frac{r_{\rm H}}{r} \right)^2  \nonumber \\
  &\approx 0.034 
  \left( \frac{\tau_{f}}{0.1} \right)^{-1/3}
  \left( \frac{M_{\rm c}}{ {\rm M}_{\rm E}} \right)^{2/3}
  \left( \frac{r}{10\,{\rm AU}} \right)^{-1/2}.
  \label{eq:frac_full}
\end{align}
The ratio of accreted to drifting pebbles depends only weakly on the pebble
surface density through the dominant particle size $\tau_{\rm f}$.
However, the filtering factor is a strong function of the embryo mass.
For low core masses, $f$ is very small, and cores only take up a negligible
fraction of the total pebble flux through the disc.  
Larger cores accrete more pebbles, but $f$ remains smaller than unity up to
critical/final cores masses are reached. 
We discuss the filtering factor in systems with multiple cores in
Section\,\ref{sec:multis}.

\subsection{Critical core mass}
\label{sec:Mcrit}

The embryo accretes pebbles and grows in mass, which leads to the
attraction of a gaseous envelope around the core. 
This atmosphere becomes more massive over time, but remains pressure
supported by the heat deposited from accreted solid material.  
However, for a given accretion rate there exist a point where the atmosphere is
no longer hydrostatically stable, triggering rapid gas accretion
\citep{Mizuno_1980}. 
The mass of the embryo at this point is standardly identified as the critical
core mass, which is typically of the order of $10$\,M$_{\rm E}$. 

In an accompanying paper \citep{Lambrechts_2014a}, we numerically
determine the critical core mass for the pebble accretion model. 
At face value, the high pebble accretion rates (Eq.\,\ref{eq:Mcdot_const}),
lead to critical core masses on the order of $100$\,M$_{\rm E}$. 
However, there exists a pebble isolation mass, a mass where the core perturbs
the gas disc and halts pebble accretion abruptly \citep{Morbidelli_2012}.
Consequently, after pebble isolation, the core is super-critical. The envelope
is no longer supported by accretional heat, the thermal balance is broken
and gas is accreted at a high rate. 
The pebble isolation mass is the result of the gravity of the core locally
perturbing the gas density, in a process similar to gap formation. In these
pressure bumps, locally the pressure gradient, and thus $\eta$, becomes zero,
stopping the drift of pebbles (Eq.\,\ref{eq:drift_ts}). 
The mass necessary to cause this perturbation and halt the accretion of pebbles
by a core is
\begin{align}
  M_{\rm iso} \approx 20 \left( \frac{r}{ 5 {\rm AU}} \right)^{3/4} 
  {\rm M}_{\rm E}\, ,
  \label{eq:iso}
\end{align}
as determined from hydrodynamical simulations \citep{Lambrechts_2014a}.
Thus, when $M_{\rm iso}$ is reached we also reach the critical core mass.
If planets do not grow past this mass, $M <M_{\rm iso}$, this typically
indicates that the planet does not become critical, leading to the formation of
an ice giant planet (a planet with a gaseous envelope much less massive than the
core).
An exception to this are planets in very wide orbits ($\gtrsim 50$\,AU), where
the critical core mass can be reached before pebble isolation (see
Section\,\ref{sec:HR8799}, Fig.\,\ref{fig:M_HR}).

\subsection{Core growth}
\label{sec:core_growth}

\begin{figure}
  \includegraphics{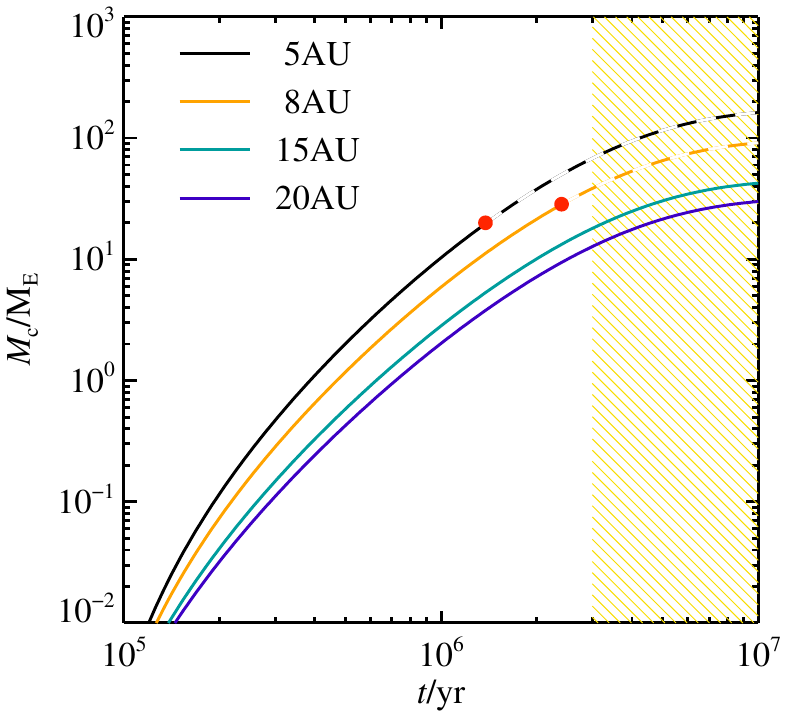}
  \caption{
  Core growth as function of time. 
  The different orbital separations ($5$-$8$-$15$-$20$\,AU) resemble the
  compact orbital configuration expected after disc dissipation
  \citep{Tsiganis_2005}. 
  The red points identify the critical mass where a phase of rapid gas
  accretion is triggered necessary for the formation of gas giants 
  \citep{Lambrechts_2014a}. 
  Core growth comes to a halt after reaching the critical core mass. 
  The dashed lines represent an extrapolation ignoring this halt in accretion.
  The yellow shaded area marks the dissipation of the gas disc after $\tau_{\rm
  dis}=3$\,Myr.
  }
  \label{fig:M}
\end{figure}

The mass of the core depends weakly on the orbital distance, but strongly on the
metallicity and surface density.
Given the pebble accretion rates, we can find the mass of the core as function
of time.
Integration of Eq.\,(\ref{eq:Mcdot_const}) yields
\begin{align}
  M_{\rm c}(t) 
  =& \left( c r^{-5/12} 
  \left[ t^{13/18} \right]^{t}_{t_{\rm i}} 
  +M_{\rm c, 0}^{1/3} \right)^{3} \nonumber\\
  \approx& 
  11
  \left( \frac{Z_0}{0.01} \right)^{25/6} 
  \left( \frac{M_*}{ {\rm M}_\odot} \right)^{-11/12}
  \left( \frac{\beta_0}{500\,{\rm g\,cm}^{-2}} \right)^{3}
  &  \nonumber \\
  &\times \left( \frac{r}{10\,{\rm AU}} \right)^{-5/4}
  \left( \frac{t}{10^6\,{\rm yr}} \right)^{13/6}
  \,{ {\rm M}_{\rm E}\,,}
  \label{eq:Mt_small}
\end{align}
where $c$ is the product of the mass, time and orbital radius independent terms
and $t_{\rm i}$ the time when the seed core of mass $M_{\rm c, 0}$ is introduced in the disc.
Here we have assumed a surface density which does not evolve in
time\footnote{The solution with exponential disc dissipation, which takes the
form of an incomplete gamma function, is calculated numerically for all results
illustrated in Fig.\,\ref{fig:M} and following similar figures.}.
The second line of Eq.\,(\ref{eq:Mt_small}) is a valid approximation in the
limit where the core is much larger than the initial seed mass $M_{\rm c}(t)
\gg M_{\rm c,0}$, at a time later than $t \gg t_{\rm i}$ in a disc with
constant surface density. 

Figure\,\ref{fig:M} illustrates how the growth of the core depends little on the
separation from the host star.
The initial embryo masses were taken to be $10^{-3}$\,M$_{\rm E}$ and inserted
at a time $t_{\rm i}=10^5$\,yr. 
Embryo growth depends little on these assumptions\footnote{
However, for consistency one has to verify that $t_{\rm i}$ is chosen such that
the pebble production line $r_{\rm g}$ has passed the orbit of the planet $r$,
so the seed planetesimal can form by the streaming instability and accrete
pebbles.
}, as can be seen in
Eq.\,(\ref{eq:Mt_small}).
The model parameters are the metallicity $Z=0.01$ and the initial gas surface
density of $500$\,g/cm$^2$ at 1\,AU.
Planets within approximately 10\,AU reach the critical core mass and trigger
rapid gas accretion (the red dot marks the pebble isolation mass), while
planets at wider orbits, which do not reach $M_{\rm iso}$, are stranded as ice giants. 

\begin{figure}
  \includegraphics{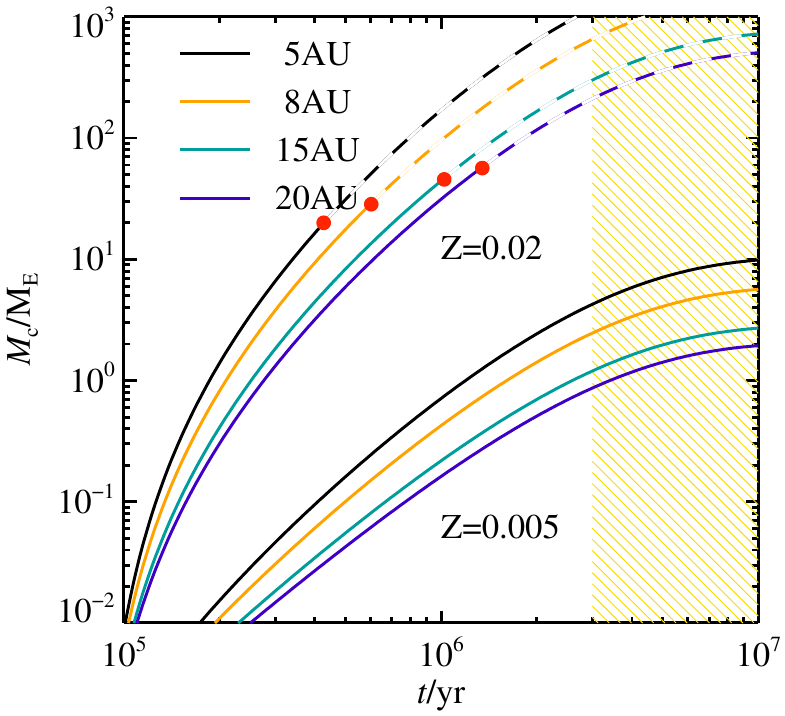}
  \caption{
  Core growth as function of time, for different values of the initial
  dust-to-gas ratio $Z_0$. Cores are placed on the same orbits as in
  Fig.\,\ref{fig:M}, and similar labeling is used. 
  Core growth is very sensitive to the initial metallicity: a twice as
  high value as the canonical dust-to-gas ratio of $Z_0=0.01$ leads to the
  formation of exclusively gas giants, while lowering the metallicity by a
  factor 2 leads to systems of small ice giants.
  }
  \label{fig:M_Zvar}
\end{figure}

\begin{figure}
  \includegraphics{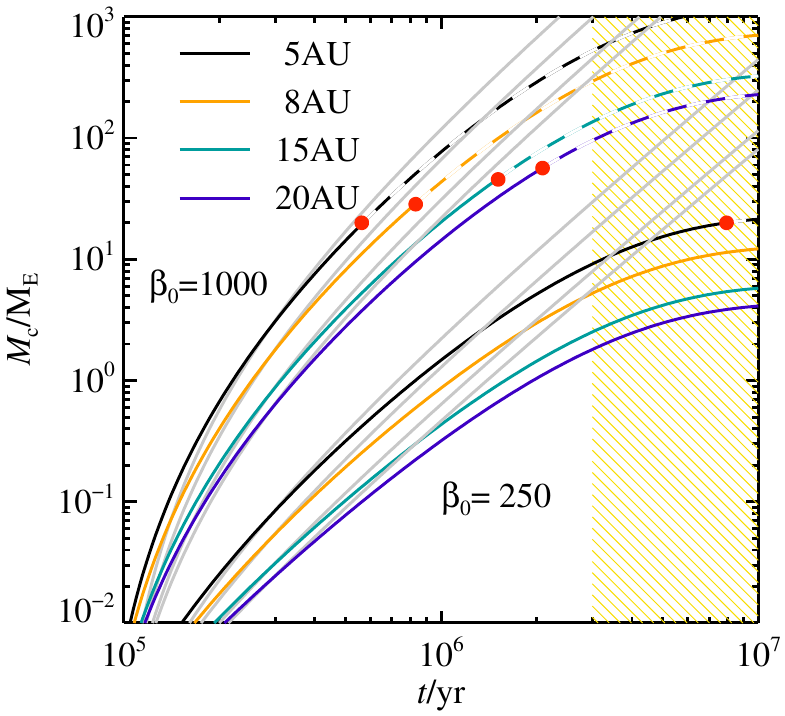}
  \caption{
  Core growth as function of time, for two values of the initial gas surface
  density $\beta_0$, which has been altered by a factor of two from the standard
  value used here ($\beta_0 = 500$\,g/cm$^2$). Cores are placed on the same
  orbits as in Fig.\,\ref{fig:M}, and similar labeling is used.
  The grey lines give the evolution in a disc with a temporally constant gas
  surface density profile, corresponding to Eq.\,(\ref{eq:Mt_small}).
  }
  \label{fig:M_beta}
\end{figure}

Core growth is highly sensitive to the metallicity, as can be seen from
Eq.\,(\ref{eq:Mt_small}). 
Figure\,\ref{fig:M_Zvar} shows the difference between an initial dust-to-gas
ratio of $Z=0.005$ and $Z=0.02$ (while keeping other parameters fixed).
The evolution of the core mass is similarly sensitive to the choice of the
(initial) gas surface density. 
Figure\,\ref{fig:M_beta} illustrates the growth of planetary cores for a gas
surface density half and double that of our standard choice ($\beta_0 =
500$g/cm$^2$), both for the cases with exponential gas dissipation over time and
without.
We discuss the sensitivity of planetary growth to the metallicity and gas
surface density in more detail in Section \ref{sec:exos}.

\section{Planetary migration}

We have assumed that cores grow approximately in situ. However,
while the core grows to embryo size, it is susceptible to Type-I migration. Due
to a torque asymmetry, the planet migrates relative to the disc towards the
star \citep{Goldreich_1980, Ward_1997}. The rate of this migration can be expressed as:
\begin{align}
  \frac{dr}{dt}  = -
  c
  \frac{M_{\rm c}}{M_*}
  \frac{\Sigma_{\rm g} r^2}{M_*}
  \left( \frac{H}{r} \right)^{-2} v_{\rm K}\,.
  \label{eq:tanaka}
\end{align}
Here, $c$ is a parameter that depends on the radial pressure and temperature
structure of the protoplanetary disc. 
\citet{Kretke_2012} give an overview of the migration rates in power-law discs
\citep{Tanaka_2002,Paardekooper_2010,Paardekooper_2011}. We adopt
$c=2.8$ in the isothermal regime \citep{Paardekooper_2010}, but other
prescriptions would only weakly change migration rates by order unity for our
simple disc model. 

By combining the planetary accretion rate (Eq.\,\ref{eq:Mcdot}) and the
migration rate, we can find the relation between the planetary mass and the
migrated distance,
\begin{align}
  \frac{dM_{\rm c}}{dr} &=
  \dot M_{\rm c} \left(\frac{dr}{dt} \right)^{-1}=
   - \frac{\kappa}{c} \gamma^2 G^{-1/12} M_*^{5/4} t^{-1/6}
   r^{-1/4} M_{\rm c}^{-1/3} \nonumber \\
   &= - K r^{-1/4} M_{\rm c}^{-1/3}\,.
  \label{eq:dmdr}
\end{align}
This expression no longer depends on $\beta$, and thus the gas surface density.
We have assumed here $H/r = \gamma r^{1/4}$, with
$\gamma=0.033$\,AU$^{-1/4}$, and defined $\kappa =
2^{11/6}3^{-5/4} \epsilon_{\rm d}^{1/3}\epsilon_{\rm p}^{-1/2}Z_0^{5/6}
(\tau_{\rm f}/0.1)^{2/3}$. 
To ease the calculation, we also fixed the pebble surface density, which only
slowly changes with time, to the profile at $t=10^6$\,yr and assumed a constant
pebble size of $\tau_{\rm f}=0.05$. The constant $K$ has the value of
$K=16$\,M$_{\rm E}^{4/3}$\,AU$^{-3/4}$.
After integration we find the planetary mass as a function of the migrated
distance,
\begin{align}
  M_{\rm c} &= \left( -\frac{16}{9}K \left( r^{3/4} - r_0^{3/4} \right)+
  M_0^{4/3}
  \right)^{3/4}\, .
  \label{eq:Masrandlim}
\end{align}
Here, $r_0$ and $M_{0}$ are the initial orbital radius and mass of the embryo.
This expression allows us to identify the lower mass above which a planetary
core rapidly moves towards the star
\begin{align}
  M_{\rm c}^\dagger &\approx  26
  \left( \frac{\tau_{\rm f}}{0.05} \right)^{8/9}
  \left( \frac{Z_0}{0.01} \right)^{5/8} 
  \left( \frac{r_0}{10\,{\rm AU}} \right)^{9/16}
  {\rm M}_{\rm E}\,.
  \label{eq:Mupp}
\end{align}
This mass is in the regime where standard type I migration prescription no
longer holds, and slower type II migration is expected to take over
\citep{Ward_1997}. 
Furthermore, around these masses, cores become critical and transition to a
phase of rapid growth in mass through gas accretion. 
Therefore, our planetary cores are not expected to migrate catastrophically
into the star, unless migration is faster than prescribed here, by a factor
$\gtrsim 3$, or pebbles are very small.

\begin{figure}
  \includegraphics{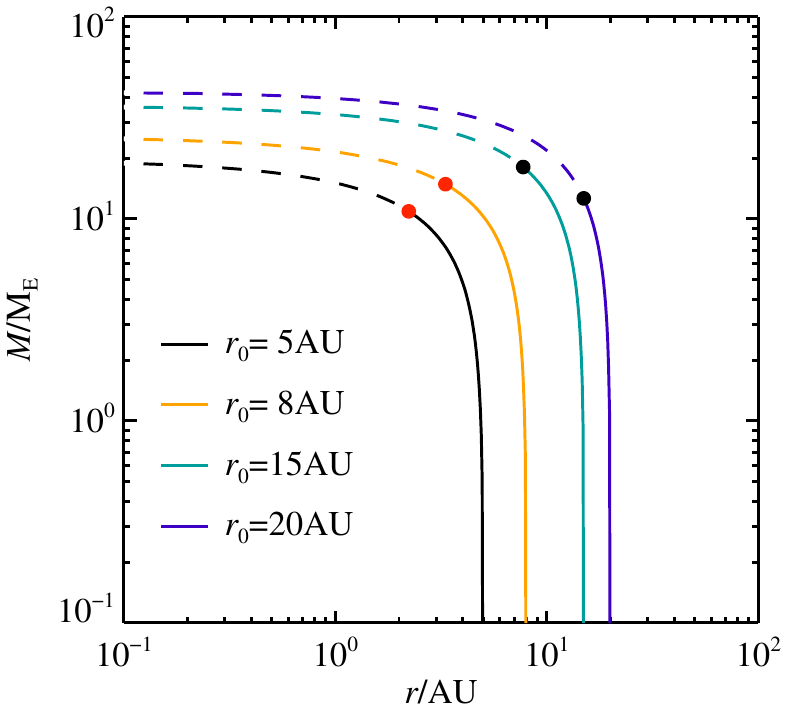}
  \caption{
  Migration of cores due to type I migration, during the pebble accretion
  phase. 
  Cores are placed initially at different initial orbital radii ($r_0 =
  5,8,15,20$\,AU). 
  Red dots mark the critical core mass and black dots the time at which the
  disc dissipates. 
  Overall, migration is modest in discs where cores grow rapidly by pebble
  accretion.
  }
  \label{fig:mvsr}
\end{figure}

An extended discussion on planetary migration is beyond the scope of this paper.
We can however consider cores in orbits similar to the early Solar
System.  
Figure\,\ref{fig:mvsr} shows the migration trajectories, by giving the
planetary mass as function of orbital radius. 
The inner planet, initially placed at 5\,AU (with a mass of $10^{-3}$\,M$_{\rm
E}$) drifts inwards for about 2\,AU, before reaching the critical core mass
(here calculated taking the migration of the core into account). 
By triggering rapid gas accretion, the core leaves the type I migration regime.  
Here we have not taken into account that the disc structure around the location
of Jupiter could significantly deviate from the simple power-law disc assumed
here. 
In such regions where the viscosity or the opacity sharply changes,
the migration of the giant planet cores can come to a complete
halt \citep{Masset_2006,Lyra_2010,Kretke_2012,Bitsch_2013,Pierens_2013}.
Our Saturn-like analogue similarly drifts a modest distance before
reaching the critical core mass. Again, convergence zones could change this
picture somewhat, and also the earlier growth of a Jupiter-like planet could
halt the migration of the Saturn-analogue.

For the ice giants we find that cores in wide orbits undergo similarly little
migration, because of the rapid core growth by pebble accretion.
As opposed to the gas giants, ice giants cannot benefit from convergence zones,
because they are generally not found to occur in the outer disc
\citep{Bitsch_2014}. 
Therefore, for planets in wider orbits growth really must be as fast as in the
pebble accretion scenario in order to prevent the embryos from being lost to
regions closer to the star.

\section{Planetary systems}
\label{sec:multis}

\begin{figure}
  \includegraphics{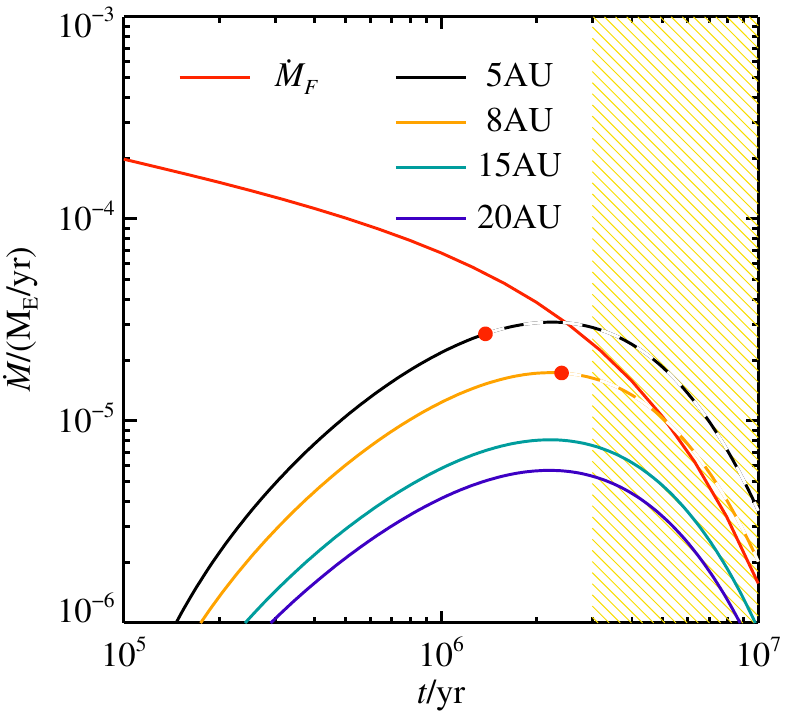}
  \caption{
  Accretion rates of planetary cores at different orbital distances and the
  mass flux in pebbles past the cores (red curve). The cores of the gas
  giants reach their critical core mass before they achieve complete filtering
  (5-8\,AU curves). 
  In wider orbits (15-20\,AU curves) planets only accrete a minor fraction of
  the full pebble flux, because the gas and dust disc dissipate after
  $\tau_{\rm dis} =3$\,Myr (yellow dashed area), well before sizes with
  efficient filtering are reached.
  }
  \label{fig:Mdot}
\end{figure}

\subsection{Dust disc mass}

Based on the growth of the innermost core, we can estimate a minimally
required mass of solids in the disc in order to form giant planets. 
As illustrated in Fig.\,\ref{fig:M} growth occurs inside-out, and therefore the
pebble flux is not significantly reduced when multiple cores are considered.
Figure\,\ref{fig:Mdot} illustrates the accretion rates of the cores, as a
function of time, for planets in the same orbits as in Fig.\,\ref{fig:M}.
For a limited number of embryos that grow into planets there is no real
`competition' for pebbles.
Thus, Fig.\,\ref{fig:M} which does not include reduction in the pebble flux on
the inner planets from the accretion of pebbles by the outer planets is also a
good description of a system of multiple planets.

We can estimate the total mass of pebbles necessary to drift towards the core,
$M_{\mathcal F}$, in order for a planet to grow to the critical core mass.
To increase the core in mass by an amount $dM_{\rm c}$, we need a mass of pebbles
$dM_{\mathcal F} = f^{-1} dM_{\rm c}$. After integration and using the filtering
factor $f$ from Eq.\,(\ref{eq:frac_full}), we find
\begin{align}
    M_{\mathcal F} \approx 130
    \left( \frac{M_{\rm c}}{20\,{\rm M}_{\rm E}} \right)^{1/3}
    \left( \frac{r}{\rm 5 AU} \right)^{1/2}
    \left( \frac{\tau_{\rm f}}{0.05} \right)^{1/3} \, {\rm M}_{\rm E}\,.
  \label{eq:minsolmass}
\end{align}
We have taken here $M_{\rm c} = 20$\,M$_{\rm E}$ for the final core mass,
corresponding to the critical core mass at $5$\,AU (Section\,\ref{sec:Mcrit}).
The efficiency of converting pebbles into a single core is thus on the order
of $M_{\rm c}/M_{\mathcal F} \approx$\,$15$\%. 

The minimum solid mass estimated in Eq.\,(\ref{eq:minsolmass}) does not increase
significantly when considering multiple cores, because of the inside-out
formation we propose.
This also implies that if there is sufficient material available in the disc to
form one core, then there is enough mass to form several more.
Therefore, planetary systems formed by pebble accretion likely consist
of multiple gas and ice-giant planets.

We briefly comment that in planetesimal-driven scenarios \citep{Pollack_1996},
it is difficult to estimate the initial total mass in solids required to form
the cores of giant planets, because the conversion efficiency of dust to
planetesimals is poorly known. 
In contrast, for core growth by pebble accretion, we can assume that pebbles
formed very efficiently. 
Particles of mm-sizes appear to be a robust outcome of dust coagulation,
both theoretically \citep{Brauer_2008, Birnstiel_2012}, observationally
\citep{Ricci_2010} and in the laboratory \citep{Blum_2008}.
Therefore, the estimate made in Eq.\,(\ref{eq:minsolmass}) directly ties the
initial dust mass in the protoplanetary disc to the efficiency of planet
formation by pebble accretion.

\subsection{Observed dust disc masses}

The required total dust mass estimated in Eq.\,(\ref{eq:minsolmass}) of about
130\,M$_{\rm E}$ is consistent with estimates of dust masses inferred in
protoplanetary discs, roughly between $1$-$300$\,M$_{\rm E}$
\citep{Andrews_2013, Mohanty_2013}. 
These measurements of disc masses should be interpreted with caution, as
they are in fact lower limits \citep{Hartmann_2008}.  
For example, particles could grow larger than mm in which case the opacity
would be overestimated or the metallicity in the disc could decrease over time,
as we see in our model (Fig.\,\ref{fig:sig}). 
For one of the best characterized protoplanetary discs, found around the young
star TW Hya, the gas mass is measured to be $M_{\rm gas} \gtrsim
0.05$\,M$_\odot$\citep{Bergin_2013}. For $Z=0.01$, this would result in a dust
disc mass of more than $167$\,M$_{\rm E}$.

The total dust mass placed in our disc model is comparable to what would be
obtained from the MMSN integrated to $100$\,AU ($\approx$$138$\,M$_{\rm E}$). 
However, by construction the MMSN does not take into account the solids that
were not accreted onto the planets.
The required total dust mass in our model remains significantly lower than the
solid surface densities enriched 4 to 6 times with respect to the MMSN required
in core growth models with planetesimals \citep{Pollack_1996}, which
additionally assume perfect dust to planetesimal conversion.  
Recent work even suggests that dust enhancements by more than a factor 10 are
necessary to explain growth of the cores of the gas giants by planetesimals
\citep{Kobayashi_2011}.

\subsection{Systems similar to the Solar System}

In Fig.\,\ref{fig:M}, we showed the growth of cores placed at $5$-$20$\,AU,
similar to the compact configuration of the solar system giant planets expected
after disc dissipation \citep{Tsiganis_2005}.
These results should be approached with caution, as we merely intend to
demonstrate the plausibility of forming planetary systems in pebble discs,
rather than to explore the \emph{exact} conditions under which the Solar System formed. 
It is nevertheless of interest to highlight some of the relatively robust
characteristics of planetary systems formed from the pebble flux in protoplanetary discs.

We first focus on the two inner cores that turn into gas giants. 
Our model naturally leads to the formation of the core of Jupiter before the
core of Saturn, and the latter typically forms close to the time of disc
dissipation.
This is a desirable feature for the Grand Tack scenario \citep{Masset_2001,
Morbidelli_2007a, Walsh_2011}, where Saturn catches up with Jupiter to share
a common gap, which leads to outward migration. This only occurs when Saturn
remains smaller than Jupiter \citep{Pierens_2011}.

The cores of the gas giant planets appear only relatively late in the disc
lifetime, after $\approx$ $1$\,Myr. 
Pebble accretion rates are reduced compared to an MMSN-based estimate, because
the instantaneous pebble column density in the evolving pebble disc is lower.
This slowdown is actually a desirable feature, as it can explain the limited gas
accretion onto Jupiter and Saturn \citep{Szul_2014} and the high noble gas
content of the gas giants \citep{Owen_1999,Guillot_2006b} as gas dissipation
increases the fraction of all condensable species.

In the outer disc cores remain small and of similar size, between $10$ and $20$\,M$_{\rm E}$.
We have verified that pebble accretion rates remain sufficiently high at all
times to prevent runaway gas accretion, unless the core reaches isolation from
pebbles (Section\,\ref{sec:Mcrit}).
In Fig.\,\ref{fig:M}, the Uranus analogue grows slightly larger than the Neptune
analogue ($18$ and $13$\,M$_{\rm E}$ respectively). 
In reality, Uranus has  $M_{\rm U}=14.5$\,M$_{\rm E}$ and Neptune has $M_{\rm
N}=17.2$\,M$_{\rm E}$.
However, here we have not included effects of gravitational perturbations
between the embryos or from the planetesimals, which could displace the planets.
Indeed, the probability that Uranus and Neptune exchanged positions after
disc dissipation is high \citep{Tsiganis_2005}.

At the time of disc dissipation (here taken to be $\approx$$3$\,Myr), of the
order of 50\,M$_{\rm E}$ of drifting pebbles are left in the giant planet
zone\footnote{
This mass in pebbles was obtained by integrating the pebble flux between the
time at which Jupiter reaches the critical core mass, halting the inwards
pebble drift, and the time of disc dissipation, while subtracting the growth of
the other three planets during that time.
}.
One can therefore imagine a scenario where during the disc dissipation phase the
metallicity is elevated by gas removal and pebbles are efficiently converted
into planetesimals by the streaming instability \citep{Johansen_2009, Bai_2010}. 
In the Nice model \citep{Tsiganis_2005}, planetesimal scattering after disc
dissipation forces the giant planets into their final orbital
architecture. To do so, a mass in planetesimals of about 50\,M$_{\rm E}$ is
required \citep{Morbidelli_2007b, Batygin_2010, Levison_2011}, in good agreement with our estimate.

It is difficult to extrapolate our results towards the inner disc, where the
terrestrial planets formed.
In our simple model, at the time a planet at $5$\,AU reaches isolation, $\approx
55$\,M$_{\rm E}$ of pebbles have drifted past the core. 
This value was calculated by integrating the pebble flux between $t=10^5$\,yr
and the time at which the Jupiter analogue reaches its critical mass, and
subtracting the mass in the cores at that time.
After these pebbles cross the ice line, about $27$\,M$_{\rm E}$ of solids are
deposited in the terrestrial planet zone, assuming a rock-to-ice fraction of 50\,\%.
This is large compared to the total mass required to form the terrestrial
planets, which is typically between $2.5$-$10$\,M$_{\rm E}$
\citep{Raymond_2013}.
However, the drift of pebbles could be halted by a pressure bump at the ice
line \citep{Kretke_2007} or reduced by the rapid growth of solids through
a sublimation-condensation cycle \citep{Ros_2013}.
However, for planetary systems different than the Solar System, emptying a reservoir of small solids into the inner planetary system
has previously been proposed to explain the rich systems of small planets
detected by the Kepler satellite \citep{Lissauer_2011}.

In summary, pebble accretion leads to inside-out planet formation, where cores
in the inner disc become gas giants, and cores in the outer disc become ice giants.
The gas giants form at a time close to disc dissipation, and with a favourable
mass ratio for a late Grand Tack migration. 
The amount of remnant pebbles converted to planetesimals is sufficient for
planetesimal driven planet migration as envisioned in the Nice model.

\subsection{Exoplanetary systems}
\label{sec:exos}

\subsubsection{Metallicity}
Core growth is very sensitive to the initially assumed metallicity $Z_0$
(Eq.\,\ref{eq:Mt_small}). 
In Fig.\,\ref{fig:M_Zvar} we demonstrated the effect of an increase and decrease
of the initial dust-to-gas ratio by a factor 2. 
In systems with a low initial dust content, below solar values, the
formation of gas giants is suppressed. 
Such a sharp cut-off could explain why exoplanet surveys find that, at least
for close in exoplanets, gas giants are nearly absent around stars with
sub-solar metallicities \citep{Fischer_2005}. Neptune-mass
planets and super-Earths, on the other hand, appear around both low- and
high-metallicity stars \citep{Buchhave_2012}.

\subsubsection{Dependence on disc and stellar mass}

\begin{figure}
  \includegraphics{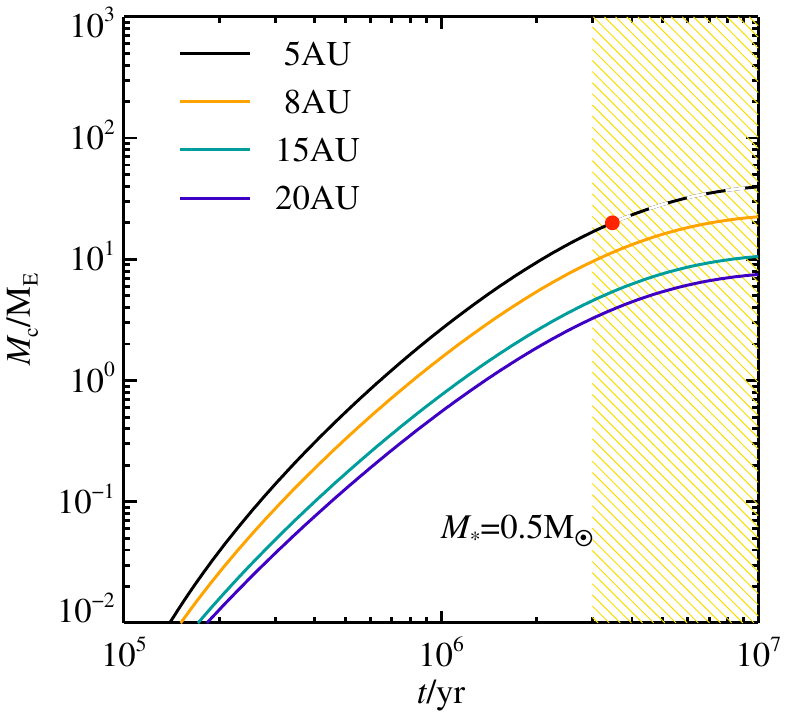}
  \caption{
  Core growth as function of time, for a low mass star. Cores are placed on the
  same orbits as in Fig.\,\ref{fig:M}, and similar labeling is used. Here we
  have assumed that a star with half the mass of the Sun has a circumstellar
  disc half as massive than considered in our solar case. The formation of
  giant planets that reach the critical core mass and become gas giants becomes
  much harder.
  }
  \label{fig:Mdot_lowmassstar}
\end{figure}

We can also explore the formation of cores in systems where we decrease the
stellar mass and the disc mass, which are observed to be approximately
proportional to each other \citep{Mohanty_2013, Andrews_2013}. 
For example, Fig.\,\ref{fig:Mdot_lowmassstar} shows the growth of cores when
the stellar and disc mass are half that of the standard values used here
($M_*=0.5M_\odot$ and an initial surface density of $250$\,g/cm$^2$ at $1$\,AU). 
No gas giant planets are formed under these conditions. 
This also seems to be supported by radial velocity and lensing surveys. Low
mass stars ($\approx$$0.5$\,M$_\odot$) rarely host gas giants
($M>100$\,M$_{\rm E}$), compared to ice giants which are found 10 times as
often, in orbits between $2.3$ and $7.2$\,AU \citep{Clanton_2014}.
Also planetary candidates in the Kepler satellite data, found in close orbits,
show a similar trend where stars with masses below $\approx 0.8$\,M$_\odot$
rarely host gas giant planets \citep{Wu_2013}.

\subsubsection{The planetary system around HR8799}
\label{sec:HR8799}

\begin{figure}
  \includegraphics{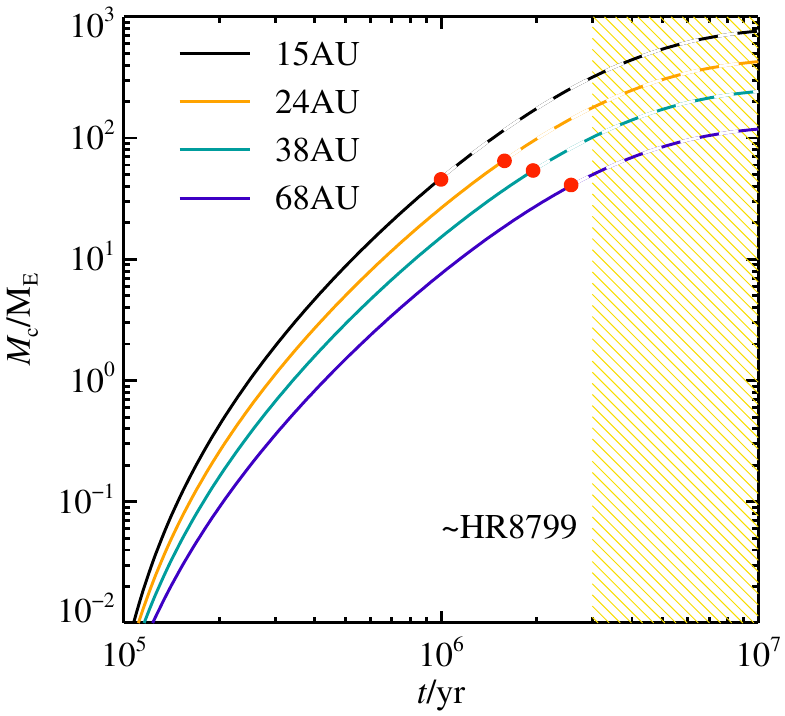}
  \caption{
  Core growth as function of time, for a system with orbital distances taken
  from the current orbits of the known exoplanets in HR8799. 
  In order to comfortably form the critical core at the widest separation of
  $68$\,AU, we increased the disc mass by a factor three compared to the Solar
  System case depicted in Fig.\,\ref{fig:M}. 
  The outer two cores reach the critical core mass before reaching pebble
  isolation (see also Section\,\ref{sec:Mcrit}).
  }
  \label{fig:M_HR}
\end{figure}

The planetary system around HR8799 consists of at least 4 planets in wide
orbits between $15$ and $70$\,AU \citep{Marois_2010}. 
In situ formation is hard to explain in the context of planetesimal-driven
core accretion theory, because planetesimal accretion rates are too slow in wide
orbits \citep{Dodson_2009}.
Although likely a rare system \citep{Lafreniere_2007}, such planets in wide
orbits can be formed in the pebble accretion scenario. 
For example, we modelled a planetary system similar to HR8799, illustrated in
Fig.\,\ref{fig:M_HR}. Here, we assumed $Z_0=0.01$, consistent with the
metallicity inferred for HR8799 \citep{Baines_2012} and similarly kept all
parameters the same as in our standard model, but increased the initial gas
surface density by a factor 3, corresponding to 1500\,g/cm$^2$ at $1$\,AU and
changed the mass the star to that of HR8799 ($1.47$\,M$_\odot$). 
Other choices {\rm to form the system} are also possible, for example one could
half the initial gas column density, but extend the disc lifetime to $\tau_{\rm
dis} =5$\,Myr or alter the metallicity, etc.

Alternatively, the HR8799 planets could have formed by gravitational
instabilities, during the formation of the protoplanetary disc
\citep{Helled_2010b}. 
However, planetary-mass objects are difficult too form at these orbital
distances \citep{Kratter_2010} and numerical simulations show that it is
difficult not to catastrophically disrupt multiple systems
formed by self-gravity \citep{Vorobyov_2013}.
Nevertheless, if these planets are the result of gravitational instabilities of
the disc, they would have a low, solar-like heavy element content, while our
pebble accretion model predicts a significant enrichment in heavy elements.

\section{Future directions}

We identify three areas for future work. 
Especially important is the incorporation of the H$_2$O, CO$_2$ and
CO icelines. The latter has recently been identified in the protoplanetary disc
around TW Hya \citep{Qi_2013}.
In these regions we can expect particle growth to large pebble sizes and large
pebble surface densities \citep{Ros_2013}, possibly aided by the emergence
of a pressure bump
\citep{Kretke_2007}. Therefore, core growth rates would increase significantly
near icelines. Furthermore, icelines recycle solid material that otherwise
would be lost by radial migration. 
This could reduce the total disc masses necessary to form the giant planets
even more.

A second important issue is that we have only traced those seeds that formed the
cores of the giant planets. 
However, it is possible that many more embryos emerged during the evolution of
the protoplanetary disc. 
It is uncertain how many seeds form with sizes that are large enough to enter
the Hill accretion regime. 
This depends on the initial planetesimal size distribution and pebble
accretion in the Bondi regime for such smaller planetesimals.
If a large amount of such seeds form initially, their continued evolution and
survival would depend on gravitational interactions between cores and
planetesimals combined with Type I migration \citep{Kretke_2013}.

A better understanding of the temporal evolution of the gas component
of the protoplanetary disc is warranted. 
In principle, one could introduce a full `alpha'-model in order to evolve the
gas surface density in time \citep{Lynden_1974}. However, the validity of such
models can be questioned. 
Self-similar solutions with an alpha model for the disc viscosity while
maintaining a constant temperature, such as in \citet{Hartmann_1998} should be
improved with models including more detailed heating physics that capture the
thermal profile of the disc \citep{Chambers_2009}. 
More crucially, it is by now rather well established that a large fraction of
protoplanetary discs cannot drive ideal MRI turbulence, weakening the case for
a simple alpha prescription.
Disc winds have recently emerged as the possible main driver of angular
momentum transport \citep{Turner_2014}. It will thus be necessary to re-address
the evolution of key quantities such as the gas surface density, the gas scale
height and turbulent strength $\alpha_{\rm t}$ in disc-wind models \citep[e.g.,][]{Armitage_2013}. 
Nevertheless, unless vertical field strengths show strong radial dependence,
discs evolve homologously: they decline equally on all radii and surface
densities are approximately inversely proportional to the orbital radius, in
agreement with the simple prescription used here.
Finally, the evolution of gas discs is possibly even more complicated if they
are not isolated, as is typically assumed, but fueled by ingoing gas and dust
accretion \citep{Nordlund_2014}.

\section{Summary}

A sufficiently high radial flux of pebbles has to be maintained over the
lifetime of a protoplanetary disc in order to explain the rapid growth
of the cores of the giant planets.
In this paper, we have demonstrated that such a pebble flow naturally emerges
when taking into account dust growth and radial migration of pebbles due to gas
drag. 

We first introduced a simple analytical model of particle growth and
pebble migration, which yields results in good agreement with advanced
coagulation codes \citep{Birnstiel_2012}.
We obtain the following results regarding the pebble component:
\begin{enumerate}
  \item Pebbles with radii of approximately mm-cm are present in the giant
    planet formation zone during the entire lifetime of the protoplanetary disc,
    provided that the disc is large ($\approx$$100$\,AU).
  \item The radial mass flux in pebbles is high, of the order of
    $10^{-4}$\,M$_{\rm E}$/yr, corresponding to a solid surface densities in
    pebbles of about $\Sigma_{\rm p} \approx 0.1(r/10{\rm
    AU})^{-3/4}$\,g/cm$^2$. 
\end{enumerate}

Equipped with a model for the evolution of the pebble solid surface density,
more powerful than simple MMSN estimates, we calculate the growth of embryos in
the outer protoplanetary disc where giant planets form. Our findings can be
summarised as follows:
\begin{enumerate}
  \item A single core forms rapidly, within about $1$\,Myr at $5$\,AU.
  \item The efficiency with which a core forms, here calculated as the ratio of
    accreted  pebbles over those that drift by, is approximately $20$\%.
  \item Outside 5\,AU, cores can grow sufficiently fast to avoid rapid orbital
    decay by type I migration.
\end{enumerate}

If the conditions to form an inner core are satisfied, we find that more cores
can form farther out. 
This inside-out growth occurs because outer cores emerge later and therefore do
not compete for pebbles with the inner cores. 
We investigated the behaviour of systems of multiple planets, which showed the
following characteristics:
\begin{enumerate}
  \item Systems similar to the Solar System appear readily under standard
    assumptions for the protoplanetary disc.
  \item However, core growth is very sensitive to the initial dust-to-gas
    ratio. Below the solar metallicity cores for gas giant planets can no
    longer form.
  \item Gas giant formation is also suppressed around low mass stars
    ($\lesssim$$0.5$\,M$_\odot$).
\end{enumerate}
These latter two points agree well with the findings from exoplanet surveys
\citep{Buchhave_2012, Clanton_2014}.

We investigated the system similar to the early Solar System in more detail.
There we identify the following key points:
\begin{enumerate}
  \item Jupiter forms first, within $\approx$$1$\,Myr, 
    even under conditions where the solid surface density is much lower than in
    standard core accretion scenarios with planetesimals.
  \item Saturn forms later (at $\approx$$2.5$\,Myr) explaining the
    smaller mass of Saturn necessary for outwards migration of Jupiter and
    Saturn.
  \item After the formation of the giant planets, a sufficient amount of solids
    is left for a planetesimal disc which can drive the planetary instability
    described in the Nice scenario.
  \item The solid mass delivered to the terrestrial region is
    sufficiently large to form the terrestrial planets.
\end{enumerate}

In summary, we have demonstrated that the pebble column density in typical
protoplanetary discs is sufficiently high to form the cores of giant planets by
pebble accretion within disc lifetimes ($\approx$$3$\,Myr), even in wide orbits
outside $10$\,AU.
Additionally, such fast core growth likely overcomes inward migration by disc torques (known as type I migration).
The efficiency with which embryos grow from the radial pebble flux is high.
Therefore, an initial mass of about $150$\,M$_{\rm E}$ in dust is
sufficient and in agreement with observations of protoplanetary discs.
Finally, in our model, gas-giant planets do not easily grow in low-mass discs
and usually do not form in discs with dust-to-gas ratios below solar.
This is in agreement with results from exoplanet surveys.

\begin{acknowledgements}
  M.L.\,thanks Alessandro Morbidelli, Akimasa Kataoka, Katrin Ros, Katherine
  Kretke, Bertram Bitsch, Seth Jacobson, Tristan Guillot and Til Birnstiel
  for valuable comments. The authors are grateful for the constructive
  feedback by an anonymous referee.
  A.J.\,and M.L.\,thank the Royal Swedish Academy of Sciences and the Knut and
  Alice Wallenberg Foundation for their financial support.
  A.J.\,was also supported by the Swedish Research Council (grant 2010-3710)
  and the European Research Council (ERC Starting Grant 278675-PEBBLE2PLANET).
\end{acknowledgements}

\bibliographystyle{aa}        
\bibliography{references_sup} 

\begin{thebibliography}{111}
\expandafter\ifx\csname natexlab\endcsname\relax\def\natexlab#1{#1}\fi

\bibitem[{{Andrews} {et~al.}(2013){Andrews}, {Rosenfeld}, {Kraus}, \&
  {Wilner}}]{Andrews_2013}
{Andrews}, S.~M., {Rosenfeld}, K.~A., {Kraus}, A.~L., \& {Wilner}, D.~J. 2013,
  \apj, 771, 129

\bibitem[{{Andrews} {et~al.}(2012){Andrews}, {Wilner}, {Hughes}, {Qi},
  {Rosenfeld}, {{\"O}berg}, {Birnstiel}, {Espaillat}, {Cieza}, {Williams},
  {Lin}, \& {Ho}}]{Andrews_2012}
{Andrews}, S.~M., {Wilner}, D.~J., {Hughes}, A.~M., {et~al.} 2012, \apj, 744,
  162

\bibitem[{{Armitage} {et~al.}(2013){Armitage}, {Simon}, \&
  {Martin}}]{Armitage_2013}
{Armitage}, P.~J., {Simon}, J.~B., \& {Martin}, R.~G. 2013, \apjl, 778, L14

\bibitem[{{Bai} \& {Stone}(2010)}]{Bai_2010}
{Bai}, X.-N. \& {Stone}, J.~M. 2010, \apj, 722, 1437

\bibitem[{{Baines} {et~al.}(2012){Baines}, {White}, {Huber}, {Jones},
  {Boyajian}, {McAlister}, {ten Brummelaar}, {Turner}, {Sturmann}, {Sturmann},
  {Goldfinger}, {Farrington}, {Riedel}, {Ireland}, {von Braun}, \&
  {Ridgway}}]{Baines_2012}
{Baines}, E.~K., {White}, R.~J., {Huber}, D., {et~al.} 2012, \apj, 761, 57

\bibitem[{{Batygin} \& {Brown}(2010)}]{Batygin_2010}
{Batygin}, K. \& {Brown}, M.~E. 2010, \apj, 716, 1323

\bibitem[{{Bergin} {et~al.}(2013){Bergin}, {Cleeves}, {Gorti}, {Zhang},
  {Blake}, {Green}, {Andrews}, {Evans}, {Henning}, {{\"O}berg}, {Pontoppidan},
  {Qi}, {Salyk}, \& {van Dishoeck}}]{Bergin_2013}
{Bergin}, E.~A., {Cleeves}, L.~I., {Gorti}, U., {et~al.} 2013, \nat, 493, 644

\bibitem[{{Birnstiel} {et~al.}(2012){Birnstiel}, {Klahr}, \&
  {Ercolano}}]{Birnstiel_2012}
{Birnstiel}, T., {Klahr}, H., \& {Ercolano}, B. 2012, \aap, 539, A148

\bibitem[{{Bitsch} {et~al.}(2013){Bitsch}, {Crida}, {Morbidelli}, {Kley}, \&
  {Dobbs-Dixon}}]{Bitsch_2013}
{Bitsch}, B., {Crida}, A., {Morbidelli}, A., {Kley}, W., \& {Dobbs-Dixon}, I.
  2013, \aap, 549, A124

\bibitem[{{Bitsch} {et~al.}(2014){Bitsch}, {Morbidelli}, {Lega}, \&
  {Crida}}]{Bitsch_2014}
{Bitsch}, B., {Morbidelli}, A., {Lega}, E., \& {Crida}, A. 2014, \aap, 564,
  A135

\bibitem[{{Blum} \& {Wurm}(2008)}]{Blum_2008}
{Blum}, J. \& {Wurm}, G. 2008, \araa, 46, 21

\bibitem[{{Brauer} {et~al.}(2008){Brauer}, {Dullemond}, \&
  {Henning}}]{Brauer_2008}
{Brauer}, F., {Dullemond}, C.~P., \& {Henning}, T. 2008, \aap, 480, 859

\bibitem[{{Buchhave} {et~al.}(2012){Buchhave}, {Latham}, {Johansen},
  {Bizzarro}, {Torres}, {Rowe}, {Batalha}, {Borucki}, {Brugamyer}, {Caldwell},
  {Bryson}, {Ciardi}, {Cochran}, {Endl}, {Esquerdo}, {Ford}, {Geary},
  {Gilliland}, {Hansen}, {Isaacson}, {Laird}, {Lucas}, {Marcy}, {Morse},
  {Robertson}, {Shporer}, {Stefanik}, {Still}, \& {Quinn}}]{Buchhave_2012}
{Buchhave}, L.~A., {Latham}, D.~W., {Johansen}, A., {et~al.} 2012, \nat, 486,
  375

\bibitem[{{Chambers}(2009)}]{Chambers_2009}
{Chambers}, J.~E. 2009, \apj, 705, 1206

\bibitem[{{Chambers}(2014)}]{Chambers_2014}
{Chambers}, J.~E. 2014, \icarus, 233, 83

\bibitem[{{Clanton} \& {Gaudi}(2014)}]{Clanton_2014}
{Clanton}, C. \& {Gaudi}, B.~S. 2014, \apj, 791, 91

\bibitem[{{Dent} {et~al.}(2005){Dent}, {Greaves}, \& {Coulson}}]{Dent_2005}
{Dent}, W.~R.~F., {Greaves}, J.~S., \& {Coulson}, I.~M. 2005, \mnras, 359, 663

\bibitem[{{Dodson-Robinson} {et~al.}(2009){Dodson-Robinson}, {Veras}, {Ford},
  \& {Beichman}}]{Dodson_2009}
{Dodson-Robinson}, S.~E., {Veras}, D., {Ford}, E.~B., \& {Beichman}, C.~A.
  2009, \apj, 707, 79

\bibitem[{{Draine}(2006)}]{Draine_2006}
{Draine}, B.~T. 2006, \apj, 636, 1114

\bibitem[{{Draine} {et~al.}(2007){Draine}, {Dale}, {Bendo}, {Gordon}, {Smith},
  {Armus}, {Engelbracht}, {Helou}, {Kennicutt}, {Li}, {Roussel}, {Walter},
  {Calzetti}, {Moustakas}, {Murphy}, {Rieke}, {Bot}, {Hollenbach}, {Sheth}, \&
  {Teplitz}}]{Draine_2007}
{Draine}, B.~T., {Dale}, D.~A., {Bendo}, G., {et~al.} 2007, \apj, 663, 866

\bibitem[{{Fischer} \& {Valenti}(2005)}]{Fischer_2005}
{Fischer}, D.~A. \& {Valenti}, J. 2005, \apj, 622, 1102

\bibitem[{{Garaud}(2007)}]{Garaud_2007}
{Garaud}, P. 2007, \apj, 671, 2091

\bibitem[{{Goldreich} \& {Tremaine}(1980)}]{Goldreich_1980}
{Goldreich}, P. \& {Tremaine}, S. 1980, \apj, 241, 425

\bibitem[{{Guillot}(2005)}]{Guillot_2005}
{Guillot}, T. 2005, Annual Review of Earth and Planetary Sciences, 33, 493

\bibitem[{{Guillot} \& {Hueso}(2006)}]{Guillot_2006b}
{Guillot}, T. \& {Hueso}, R. 2006, \mnras, 367, L47

\bibitem[{{Guillot} {et~al.}(2014){Guillot}, {Ida}, \& {Ormel}}]{Guillot_2014}
{Guillot}, T., {Ida}, S., \& {Ormel}, C.~W. 2014, ArXiv e-prints

\bibitem[{{Guillot} {et~al.}(2006){Guillot}, {Santos}, {Pont}, {Iro}, {Melo},
  \& {Ribas}}]{Guillot_2006}
{Guillot}, T., {Santos}, N.~C., {Pont}, F., {et~al.} 2006, \aap, 453, L21

\bibitem[{{Haisch} {et~al.}(2001){Haisch}, {Lada}, \& {Lada}}]{Haisch_2001}
{Haisch}, Jr., K.~E., {Lada}, E.~A., \& {Lada}, C.~J. 2001, \apjl, 553, L153

\bibitem[{{Hartmann}(2008)}]{Hartmann_2008}
{Hartmann}, L. 2008, Physica Scripta Volume T, 130, 014012

\bibitem[{{Hartmann} {et~al.}(1998){Hartmann}, {Calvet}, {Gullbring}, \&
  {D'Alessio}}]{Hartmann_1998}
{Hartmann}, L., {Calvet}, N., {Gullbring}, E., \& {D'Alessio}, P. 1998, \apj,
  495, 385

\bibitem[{{Hayashi}(1981)}]{Hayashi_1981}
{Hayashi}, C. 1981, Progress of Theoretical Physics Supplement, 70, 35

\bibitem[{{Helled} {et~al.}(2011){Helled}, {Anderson}, {Podolak}, \&
  {Schubert}}]{Helled_2011}
{Helled}, R., {Anderson}, J.~D., {Podolak}, M., \& {Schubert}, G. 2011, \apj,
  726, 15

\bibitem[{{Helled} \& {Bodenheimer}(2010)}]{Helled_2010b}
{Helled}, R. \& {Bodenheimer}, P. 2010, \icarus, 207, 503

\bibitem[{{Hern{\'a}ndez} {et~al.}(2005){Hern{\'a}ndez}, {Calvet}, {Hartmann},
  {Brice{\~n}o}, {Sicilia-Aguilar}, \& {Berlind}}]{Hernandez_2005}
{Hern{\'a}ndez}, J., {Calvet}, N., {Hartmann}, L., {et~al.} 2005, \aj, 129, 856

\bibitem[{{Isella} {et~al.}(2009){Isella}, {Carpenter}, \&
  {Sargent}}]{Isella_2009}
{Isella}, A., {Carpenter}, J.~M., \& {Sargent}, A.~I. 2009, \apj, 701, 260

\bibitem[{{Johansen} \& {Lacerda}(2010)}]{Johansen_2010}
{Johansen}, A. \& {Lacerda}, P. 2010, \mnras, 404, 475

\bibitem[{{Johansen} {et~al.}(2009){Johansen}, {Youdin}, \& {Mac
  Low}}]{Johansen_2009}
{Johansen}, A., {Youdin}, A., \& {Mac Low}, M.-M. 2009, \apjl, 704, L75

\bibitem[{{Johansen} {et~al.}(2012){Johansen}, {Youdin}, \&
  {Lithwick}}]{Johansen_2012}
{Johansen}, A., {Youdin}, A.~N., \& {Lithwick}, Y. 2012, \aap, 537, A125

\bibitem[{{Kataoka} {et~al.}(2013){Kataoka}, {Tanaka}, {Okuzumi}, \&
  {Wada}}]{Kataoka_2013}
{Kataoka}, A., {Tanaka}, H., {Okuzumi}, S., \& {Wada}, K. 2013, \aap, 557, L4

\bibitem[{{Kitamura} {et~al.}(2002){Kitamura}, {Momose}, {Yokogawa}, {Kawabe},
  {Tamura}, \& {Ida}}]{Kitamura_2002}
{Kitamura}, Y., {Momose}, M., {Yokogawa}, S., {et~al.} 2002, \apj, 581, 357

\bibitem[{{Kobayashi} {et~al.}(2011){Kobayashi}, {Tanaka}, \&
  {Krivov}}]{Kobayashi_2011}
{Kobayashi}, H., {Tanaka}, H., \& {Krivov}, A.~V. 2011, \apj, 738, 35

\bibitem[{{Kratter} {et~al.}(2010){Kratter}, {Murray-Clay}, \&
  {Youdin}}]{Kratter_2010}
{Kratter}, K.~M., {Murray-Clay}, R.~A., \& {Youdin}, A.~N. 2010, \apj, 710,
  1375

\bibitem[{{Kraus} {et~al.}(2012){Kraus}, {Ireland}, {Hillenbrand}, \&
  {Martinache}}]{Kraus_2012}
{Kraus}, A.~L., {Ireland}, M.~J., {Hillenbrand}, L.~A., \& {Martinache}, F.
  2012, \apj, 745, 19

\bibitem[{{Kretke} \& {Levison}(2013)}]{Kretke_2013}
{Kretke}, K.~A. \& {Levison}, H.~F. 2013, in AAS/Division for Planetary
  Sciences Meeting Abstracts, Vol.~45, AAS/Division for Planetary Sciences
  Meeting Abstracts, 415.11

\bibitem[{{Kretke} \& {Lin}(2007)}]{Kretke_2007}
{Kretke}, K.~A. \& {Lin}, D.~N.~C. 2007, \apjl, 664, L55

\bibitem[{{Kretke} \& {Lin}(2012)}]{Kretke_2012}
{Kretke}, K.~A. \& {Lin}, D.~N.~C. 2012, \apj, 755, 74

\bibitem[{{Lafreni{\`e}re} {et~al.}(2007){Lafreni{\`e}re}, {Doyon}, {Marois},
  {Nadeau}, {Oppenheimer}, {Roche}, {Rigaut}, {Graham}, {Jayawardhana},
  {Johnstone}, {Kalas}, {Macintosh}, \& {Racine}}]{Lafreniere_2007}
{Lafreni{\`e}re}, D., {Doyon}, R., {Marois}, C., {et~al.} 2007, \apj, 670, 1367

\bibitem[{{Lambrechts} \& {Johansen}(2012)}]{Lambrechts_2012}
{Lambrechts}, M. \& {Johansen}, A. 2012, \aap, 544, A32

\bibitem[{{Lambrechts} {et~al.}(2014){Lambrechts}, {Johansen}, \&
  {Morbidelli}}]{Lambrechts_2014a}
{Lambrechts}, M., {Johansen}, A., \& {Morbidelli}, A. 2014, ArXiv e-prints

\bibitem[{{Levison} {et~al.}(2011){Levison}, {Morbidelli}, {Tsiganis},
  {Nesvorn{\'y}}, \& {Gomes}}]{Levison_2011}
{Levison}, H.~F., {Morbidelli}, A., {Tsiganis}, K., {Nesvorn{\'y}}, D., \&
  {Gomes}, R. 2011, \aj, 142, 152

\bibitem[{{Levison} {et~al.}(2010){Levison}, {Thommes}, \&
  {Duncan}}]{Levison_2010}
{Levison}, H.~F., {Thommes}, E., \& {Duncan}, M.~J. 2010, \aj, 139, 1297

\bibitem[{{Lissauer} {et~al.}(2011){Lissauer}, {Fabrycky}, {Ford}, {Borucki},
  {Fressin}, {Marcy}, {Orosz}, {Rowe}, {Torres}, {Welsh}, {Batalha}, {Bryson},
  {Buchhave}, {Caldwell}, {Carter}, {Charbonneau}, {Christiansen}, {Cochran},
  {Desert}, {Dunham}, {Fanelli}, {Fortney}, {Gautier}, {Geary}, {Gilliland},
  {Haas}, {Hall}, {Holman}, {Koch}, {Latham}, {Lopez}, {McCauliff}, {Miller},
  {Morehead}, {Quintana}, {Ragozzine}, {Sasselov}, {Short}, \&
  {Steffen}}]{Lissauer_2011}
{Lissauer}, J.~J., {Fabrycky}, D.~C., {Ford}, E.~B., {et~al.} 2011, \nat, 470,
  53

\bibitem[{{Lommen} {et~al.}(2009){Lommen}, {Maddison}, {Wright}, {van
  Dishoeck}, {Wilner}, \& {Bourke}}]{Lommen_2009}
{Lommen}, D., {Maddison}, S.~T., {Wright}, C.~M., {et~al.} 2009, \aap, 495, 869

\bibitem[{{Lynden-Bell} \& {Pringle}(1974)}]{Lynden_1974}
{Lynden-Bell}, D. \& {Pringle}, J.~E. 1974, \mnras, 168, 603

\bibitem[{{Lyra} {et~al.}(2010){Lyra}, {Paardekooper}, \& {Mac
  Low}}]{Lyra_2010}
{Lyra}, W., {Paardekooper}, S.-J., \& {Mac Low}, M.-M. 2010, \apjl, 715, L68

\bibitem[{{Marois} {et~al.}(2010){Marois}, {Zuckerman}, {Konopacky},
  {Macintosh}, \& {Barman}}]{Marois_2010}
{Marois}, C., {Zuckerman}, B., {Konopacky}, Q.~M., {Macintosh}, B., \&
  {Barman}, T. 2010, \nat, 468, 1080

\bibitem[{{Masset} \& {Snellgrove}(2001)}]{Masset_2001}
{Masset}, F. \& {Snellgrove}, M. 2001, \mnras, 320, L55

\bibitem[{{Masset} {et~al.}(2006){Masset}, {Morbidelli}, {Crida}, \&
  {Ferreira}}]{Masset_2006}
{Masset}, F.~S., {Morbidelli}, A., {Crida}, A., \& {Ferreira}, J. 2006, \apj,
  642, 478

\bibitem[{{McNeil} {et~al.}(2005){McNeil}, {Duncan}, \&
  {Levison}}]{McNeil_2005}
{McNeil}, D., {Duncan}, M., \& {Levison}, H.~F. 2005, \aj, 130, 2884

\bibitem[{{Menu} {et~al.}(2014){Menu}, {van Boekel}, {Henning}, {Chandler},
  {Linz}, {Benisty}, {Lacour}, {Min}, {Waelkens}, {Andrews}, {Calvet},
  {Carpenter}, {Corder}, {Deller}, {Greaves}, {Harris}, {Isella}, {Kwon},
  {Lazio}, {Le Bouquin}, {M{\'e}nard}, {Mundy}, {P{\'e}rez}, {Ricci},
  {Sargent}, {Storm}, {Testi}, \& {Wilner}}]{Menu_2014}
{Menu}, J., {van Boekel}, R., {Henning}, T., {et~al.} 2014, \aap, 564, A93

\bibitem[{{Miller} \& {Fortney}(2011)}]{Miller_2011}
{Miller}, N. \& {Fortney}, J.~J. 2011, \apjl, 736, L29

\bibitem[{{Mizuno}(1980)}]{Mizuno_1980}
{Mizuno}, H. 1980, Progress of Theoretical Physics, 64, 544

\bibitem[{{Mohanty} {et~al.}(2013){Mohanty}, {Greaves}, {Mortlock}, {Pascucci},
  {Scholz}, {Thompson}, {Apai}, {Lodato}, \& {Looper}}]{Mohanty_2013}
{Mohanty}, S., {Greaves}, J., {Mortlock}, D., {et~al.} 2013, \apj, 773, 168

\bibitem[{{Morbidelli} {et~al.}(2009){Morbidelli}, {Bottke}, {Nesvorn{\'y}}, \&
  {Levison}}]{Morbidelli_2009}
{Morbidelli}, A., {Bottke}, W.~F., {Nesvorn{\'y}}, D., \& {Levison}, H.~F.
  2009, \icarus, 204, 558

\bibitem[{{Morbidelli} \& {Crida}(2007)}]{Morbidelli_2007a}
{Morbidelli}, A. \& {Crida}, A. 2007, \icarus, 191, 158

\bibitem[{{Morbidelli} \& {Nesvorny}(2012)}]{Morbidelli_2012}
{Morbidelli}, A. \& {Nesvorny}, D. 2012, \aap, 546, A18

\bibitem[{{Morbidelli} {et~al.}(2007){Morbidelli}, {Tsiganis}, {Crida},
  {Levison}, \& {Gomes}}]{Morbidelli_2007b}
{Morbidelli}, A., {Tsiganis}, K., {Crida}, A., {Levison}, H.~F., \& {Gomes}, R.
  2007, \aj, 134, 1790

\bibitem[{{Moutou} {et~al.}(2013){Moutou}, {Deleuil}, {Guillot}, {Baglin},
  {Bord{\'e}}, {Bouchy}, {Cabrera}, {Csizmadia}, \& {Deeg}}]{Moutou_2013}
{Moutou}, C., {Deleuil}, M., {Guillot}, T., {et~al.} 2013, \icarus, 226, 1625

\bibitem[{{Nakagawa} {et~al.}(1986){Nakagawa}, {Sekiya}, \&
  {Hayashi}}]{Nakagawa_1986}
{Nakagawa}, Y., {Sekiya}, M., \& {Hayashi}, C. 1986, \icarus, 67, 375

\bibitem[{{Natta} {et~al.}(2007){Natta}, {Testi}, {Calvet}, {Henning},
  {Waters}, \& {Wilner}}]{Natta_2007}
{Natta}, A., {Testi}, L., {Calvet}, N., {et~al.} 2007, Protostars and Planets
  V, 767

\bibitem[{{Nordlund} {et~al.}(2014){Nordlund}, {Haugb{\o}lle}, {K{\"u}ffmeier},
  {Padoan}, \& {Vasileiades}}]{Nordlund_2014}
{Nordlund}, {\AA}., {Haugb{\o}lle}, T., {K{\"u}ffmeier}, M., {Padoan}, P., \&
  {Vasileiades}, A. 2014, in IAU Symposium, Vol. 299, IAU Symposium, ed.
  M.~{Booth}, B.~C. {Matthews}, \& J.~R. {Graham}, 131--135

\bibitem[{{Okuzumi} {et~al.}(2012){Okuzumi}, {Tanaka}, {Kobayashi}, \&
  {Wada}}]{Okuzumi_2012}
{Okuzumi}, S., {Tanaka}, H., {Kobayashi}, H., \& {Wada}, K. 2012, \apj, 752,
  106

\bibitem[{{Ono} {et~al.}(2014){Ono}, {Nomura}, \& {Takeuchi}}]{Ono_2014}
{Ono}, T., {Nomura}, H., \& {Takeuchi}, T. 2014, \apj, 787, 37

\bibitem[{{Ormel} \& {Cuzzi}(2007)}]{Ormel_2007}
{Ormel}, C.~W. \& {Cuzzi}, J.~N. 2007, \aap, 466, 413

\bibitem[{{Ormel} \& {Klahr}(2010)}]{Ormel_2010}
{Ormel}, C.~W. \& {Klahr}, H.~H. 2010, \aap, 520, A43

\bibitem[{{Ormel} \& {Kobayashi}(2012)}]{Ormel_2012}
{Ormel}, C.~W. \& {Kobayashi}, H. 2012, \apj, 747, 115

\bibitem[{{Owen} {et~al.}(1999){Owen}, {Mahaffy}, {Niemann}, {Atreya},
  {Donahue}, {Bar-Nun}, \& {de Pater}}]{Owen_1999}
{Owen}, T., {Mahaffy}, P., {Niemann}, H.~B., {et~al.} 1999, \nat, 402, 269

\bibitem[{{Paardekooper} {et~al.}(2010){Paardekooper}, {Baruteau}, {Crida}, \&
  {Kley}}]{Paardekooper_2010}
{Paardekooper}, S.-J., {Baruteau}, C., {Crida}, A., \& {Kley}, W. 2010, \mnras,
  401, 1950

\bibitem[{{Paardekooper} {et~al.}(2011){Paardekooper}, {Baruteau}, \&
  {Kley}}]{Paardekooper_2011}
{Paardekooper}, S.-J., {Baruteau}, C., \& {Kley}, W. 2011, \mnras, 410, 293

\bibitem[{{P{\'e}rez} {et~al.}(2012){P{\'e}rez}, {Carpenter}, {Chandler},
  {Isella}, {Andrews}, {Ricci}, {Calvet}, {Corder}, {Deller}, {Dullemond},
  {Greaves}, {Harris}, {Henning}, {Kwon}, {Lazio}, {Linz}, {Mundy}, {Sargent},
  {Storm}, {Testi}, \& {Wilner}}]{Perez_2012}
{P{\'e}rez}, L.~M., {Carpenter}, J.~M., {Chandler}, C.~J., {et~al.} 2012,
  \apjl, 760, L17

\bibitem[{{Pierens} {et~al.}(2013){Pierens}, {Cossou}, \&
  {Raymond}}]{Pierens_2013}
{Pierens}, A., {Cossou}, C., \& {Raymond}, S.~N. 2013, \aap, 558, A105

\bibitem[{{Pierens} \& {Raymond}(2011)}]{Pierens_2011}
{Pierens}, A. \& {Raymond}, S.~N. 2011, \aap, 533, A131

\bibitem[{{Pollack} {et~al.}(1996){Pollack}, {Hubickyj}, {Bodenheimer},
  {Lissauer}, {Podolak}, \& {Greenzweig}}]{Pollack_1996}
{Pollack}, J.~B., {Hubickyj}, O., {Bodenheimer}, P., {et~al.} 1996, \icarus,
  124, 62

\bibitem[{{Qi} {et~al.}(2013){Qi}, {{\"O}berg}, {Wilner}, {D'Alessio},
  {Bergin}, {Andrews}, {Blake}, {Hogerheijde}, \& {van Dishoeck}}]{Qi_2013}
{Qi}, C., {{\"O}berg}, K.~I., {Wilner}, D.~J., {et~al.} 2013, Science, 341, 630

\bibitem[{{Rafikov}(2004)}]{Rafikov_2004}
{Rafikov}, R.~R. 2004, \aj, 128, 1348

\bibitem[{{Raymond} {et~al.}(2013){Raymond}, {Kokubo}, {Morbidelli},
  {Morishima}, \& {Walsh}}]{Raymond_2013}
{Raymond}, S.~N., {Kokubo}, E., {Morbidelli}, A., {Morishima}, R., \& {Walsh},
  K.~J. 2013, ArXiv e-prints

\bibitem[{{Ricci} {et~al.}(2010){Ricci}, {Testi}, {Natta}, {Neri}, {Cabrit}, \&
  {Herczeg}}]{Ricci_2010}
{Ricci}, L., {Testi}, L., {Natta}, A., {et~al.} 2010, \aap, 512, A15

\bibitem[{{Ros} \& {Johansen}(2013)}]{Ros_2013}
{Ros}, K. \& {Johansen}, A. 2013, \aap, 552, A137

\bibitem[{{Scaife}(2013)}]{Scaife_2013}
{Scaife}, A.~M.~M. 2013, \mnras, 435, 1139

\bibitem[{{Szul{\'a}gyi} {et~al.}(2014){Szul{\'a}gyi}, {Morbidelli}, {Crida},
  \& {Masset}}]{Szul_2014}
{Szul{\'a}gyi}, J., {Morbidelli}, A., {Crida}, A., \& {Masset}, F. 2014, \apj,
  782, 65

\bibitem[{{Tanaka} {et~al.}(2002){Tanaka}, {Takeuchi}, \& {Ward}}]{Tanaka_2002}
{Tanaka}, H., {Takeuchi}, T., \& {Ward}, W.~R. 2002, \apj, 565, 1257

\bibitem[{{Toomre}(1964)}]{Toomre_1964}
{Toomre}, A. 1964, \apj, 139, 1217

\bibitem[{{Trotta} {et~al.}(2013){Trotta}, {Testi}, {Natta}, {Isella}, \&
  {Ricci}}]{Trotta_2013}
{Trotta}, F., {Testi}, L., {Natta}, A., {Isella}, A., \& {Ricci}, L. 2013,
  \aap, 558, A64

\bibitem[{{Tsiganis} {et~al.}(2005){Tsiganis}, {Gomes}, {Morbidelli}, \&
  {Levison}}]{Tsiganis_2005}
{Tsiganis}, K., {Gomes}, R., {Morbidelli}, A., \& {Levison}, H.~F. 2005, \nat,
  435, 459

\bibitem[{{Turner} {et~al.}(2014){Turner}, {Fromang}, {Gammie}, {Klahr},
  {Lesur}, {Wardle}, \& {Bai}}]{Turner_2014}
{Turner}, N.~J., {Fromang}, S., {Gammie}, C., {et~al.} 2014, ArXiv e-prints

\bibitem[{{Ubach} {et~al.}(2012){Ubach}, {Maddison}, {Wright}, {Wilner},
  {Lommen}, \& {Koribalski}}]{Ubach_2012}
{Ubach}, C., {Maddison}, S.~T., {Wright}, C.~M., {et~al.} 2012, \mnras, 425,
  3137

\bibitem[{{Vacca} \& {Sandell}(2011)}]{Vacca_2011}
{Vacca}, W.~D. \& {Sandell}, G. 2011, \apj, 732, 8

\bibitem[{{Vorobyov}(2013)}]{Vorobyov_2013}
{Vorobyov}, E.~I. 2013, \aap, 552, A129

\bibitem[{{Walsh} {et~al.}(2011){Walsh}, {Morbidelli}, {Raymond}, {O'Brien}, \&
  {Mandell}}]{Walsh_2011}
{Walsh}, K.~J., {Morbidelli}, A., {Raymond}, S.~N., {O'Brien}, D.~P., \&
  {Mandell}, A.~M. 2011, \nat, 475, 206

\bibitem[{{Ward}(1997)}]{Ward_1997}
{Ward}, W.~R. 1997, \icarus, 126, 261

\bibitem[{{Weidenschilling}(1977)}]{Weidenschilling_1977}
{Weidenschilling}, S.~J. 1977, \mnras, 180, 57

\bibitem[{{Weidenschilling}(1984)}]{Weidenschilling_1984}
{Weidenschilling}, S.~J. 1984, \icarus, 60, 553

\bibitem[{{Weidenschilling}(2006)}]{Weidenschilling_2006}
{Weidenschilling}, S.~J. 2006, \icarus, 181, 572

\bibitem[{{Weingartner} \& {Draine}(2001)}]{Weingartner_2001}
{Weingartner}, J.~C. \& {Draine}, B.~T. 2001, \apj, 548, 296

\bibitem[{{Williams} \& {Cieza}(2011)}]{Williams_2011}
{Williams}, J.~P. \& {Cieza}, L.~A. 2011, \araa, 49, 67

\bibitem[{{Wilner} {et~al.}(2005){Wilner}, {D'Alessio}, {Calvet}, {Claussen},
  \& {Hartmann}}]{Wilner_2005}
{Wilner}, D.~J., {D'Alessio}, P., {Calvet}, N., {Claussen}, M.~J., \&
  {Hartmann}, L. 2005, \apjl, 626, L109

\bibitem[{{Windmark} {et~al.}(2012){Windmark}, {Birnstiel}, {G{\"u}ttler},
  {Blum}, {Dullemond}, \& {Henning}}]{Windmark_2012}
{Windmark}, F., {Birnstiel}, T., {G{\"u}ttler}, C., {et~al.} 2012, \aap, 540,
  A73

\bibitem[{{Wu} \& {Lithwick}(2013)}]{Wu_2013}
{Wu}, Y. \& {Lithwick}, Y. 2013, \apj, 772, 74

\bibitem[{{Wyatt}(2008)}]{Wyatt_2008}
{Wyatt}, M.~C. 2008, \araa, 46, 339

\bibitem[{{Youdin} \& {Lithwick}(2007)}]{Youdin_2007}
{Youdin}, A.~N. \& {Lithwick}, Y. 2007, \icarus, 192, 588

\bibitem[{{Youdin} \& {Shu}(2002)}]{Youdin_2002}
{Youdin}, A.~N. \& {Shu}, F.~H. 2002, \apj, 580, 494

\end{thebibliography}

\end{document}